\documentclass[a4paper,fleqn]{cas-sc}

\usepackage[authoryear]{natbib}
\usepackage{subfigure}
\usepackage{caption}

\newtheorem{theorem}{Theorem}
\newdefinition{rmk}{Remark}
\newproof{pf}{Proof}

\newcommand*{\caso}[1]{({#1})}

\begin{document}
\let\WriteBookmarks\relax
\def\floatpagepagefraction{1}
\def\textpagefraction{.001}

\shorttitle{}    

\shortauthors{}  

\title [mode = title]{Bilinearization-reduction approach to the classical and nonlocal semi-discrete modified Korteweg-de Vries equations with nonzero backgrounds}  



%

\author[1]{Xiao Deng}






\affiliation[1]{organization={Research Center for Data Hub and Security, Zhejiang Lab},
            city={Hangzhou},
            postcode={311100}, 
            state={Zhejiang Province},
            country={China}}

\author[1]{Hongyang Chen}
\author[2]{Song-Lin Zhao}





\affiliation[2]{organization={School of Mathematical Sciences, Zhejiang University of Technology},
            city={Hangzhou},
            postcode={310023}, 
            state={Zhejiang Province},
            country={China}}

\author[3]{Guanlong Ren}[orcid=0009-0001-2573-4276]
\cormark[1]
\ead{rengl@zhejianglab.org}
\affiliation[3]{organization={Research Center for Space Computing System, Zhejiang Lab},
            city={Hangzhou},
            postcode={311100}, 
            state={Zhejiang Province},
            country={China}}

\cortext[1]{Corresponding author}



\begin{abstract}
    Quasi double Casoratian solutions are derived for a bilinear system reformulated from the coupled semi-discrete modified Korteweg-de Vries equations with nonzero backgrounds. These solutions, when applied with the classical and nonlocal reduction techniques,  also satisfy the corresponding classical and nonlocal semi-discrete modified Korteweg-de Vries equations with nonzero backgrounds. They can be expressed explicitly, allowing for an easy investigation of the dynamics of systems.  As illustrative examples, the dynamics of solitonic, periodic and rational solutions with a plane wave background are examined for the focusing semi-discrete Korteweg-de Vries equation and the defocusing reverse-space-time complex semi-discrete Korteweg-de Vries equation.  
\end{abstract}


\begin{highlights}
\item The quasi double Casoratian solutions to the classical and nonlocal semi-discrete modified Korteweg-de Vries equations with nonzero backgrounds are found. 
\item Solitonic, periodic and rational solutions are obtained by using the quasi double Casoratians. 
\end{highlights}


\maketitle

\section{Introduction}\label{sec1}

This paper aims to investigate solutions with nonzero backgrounds for the (classical real) semi-discrete modified Korteweg-de Vries (sd-mKdV) equation \citep{Hirota-JPSJ1972}
\begin{eqnarray}
\partial_tQ_{n}=(1-\delta Q_nQ_n)(Q_{n+1}-Q_{n-1}) 
\end{eqnarray}
and its classical and nonlocal analogous (see eqs.\eqref{sec2-sd-mKdV-eqs1}).
Here, $Q_n=Q(n,t)$ is a function of discretized-space and time $(n,t)\in \mathbb Z\times \mathbb R$, $\partial_t$ denotes  the derivative to time $t$ and $\delta=\pm1$. This equation is termed focusing sd-mKdV when $\delta=-1$ and defocusing sd-mKdV when $\delta=1$.

The sd-mKdV equation, which can be derived from the Ablowitz-Ladik (AL) spectral problem \citep{AblowitzL-JMP1976} and  treated as a modified Volterra equation\citep{Wadati-PTPS1976}, is an important integrable spatial discretization of the mKdV equation \citep{Wadati-JPSJ1973}. The mKdV equation, which is also referred as the 3rd (reduced) equation in Ablowitz-Kaup-Newell-Segur (AKNS) hierarchy \citep{AblowitzKNS-PRL1973}, plays a vital role in mathematically modeling various nonlinear physical phenomena, such as  hydromagnetic waves in  plasma \citep{KatutaniO-JPSJ1969} and optical pulses in nonlinear media \citep{LeblondS-PRA2003,LeblondM-PR2013} (see \citet{ZhangZSZ-RMP2014} for more examples  and references therein).   
Furthermore, the sd-mKdV equation itself arises in nonlinear  models describing wave propagation in ladder type network \citep{Hirota-JPSJ1972}, motions of discrete curves \citep{DoliwaS-JMP1995} and motions of discrete surfaces \citep{HisakadoW-JPSJ1995}. 
As an integrable system, the sd-mKdV equation is also characterized with rich mathematical structures like Lax pair,  infinite many conservation laws and infinite many symmetries  \citep{AblowitzL-JMP1976,FuQSZ-JNMP2015}.


When a nonlocal version of the nonlinear Schr\"odinger (NLS) equation, which is able to describe wave phenomena in nonlinear parity-time (PT) symmetric media and admits the PT symmetry \citep{SarmaMMC-PRE2014}, is proposed by \cite{AblowitzM-PRL2013}, nonlocal integrable systems attract much attention and become a hot research topic. Like many other nonlocal integrable systems, nonlocal (real and complex) sd-mKdV equations can be obtained from a coupled system with nonlocal reductions    and inherit many characteristics  found in classical integrable systems \citep{AblowitzM-SAM2017}.

Many methods \citep{Common-IP1993,MukaihiraN-IP2000,ZhangJD-AMC2012,ZhuGK-Arxiv2013,WangM-MMAS2010,AnkiewiczAS-PRES2010,SunZ-MPLB2019,FengZS-IJMPB2020} are applied to solve the sd-mKdV equation and its analogous exactly.
Among these, the bilinearization-reduction approach proposed by \cite{ChenDLZ-SAM2018} is a powerful technique in solving the classical and nonlocal integrable systems.  This method first constructs  double Wronskian/Casoratian solutions for a coupled integrable system using associated bilinear forms, and then enforces these solutions to satisfy the classical and nonlocal reduced systems \citep{Zhang-APS2023} by directly imposing reduction constraints. This approach has been successfully applied to various integrable systems with zero background, such as AKNS hierarchy \citep{ChenDLZ-SAM2018}, negative order AKNS hierarchy \citep{WangWZ-CTP2020}, derivative NLS equations \citep{ShiSZ-ND2019} and sdNLS equations \citep{DengLZ-AMC2018} (for more examples see \cite{Zhang-APS2023} and reference therein). Recently, this method has also been implemented for 
NLS equations with nonzero backgrounds \citep{ZhangLD-OCNMP2023} and sdNLS equations with nonzero backgrounds \citep{DengCCZ-arXiv2024}. This progress motivates us to apply the bilinearization-reduction approach to integrable sd-mKdV equations with nonzero backgrounds. Here, solutions with nonzero backgrounds are characterized by the property that they do not tend to zero as $|n|\rightarrow \infty$.


This paper is organized as follows. In Sec.2, the unreduced system of the  coupled sd-mKdV equations, along with its integrable reductions are presented. Sec.3 presents the  Casoratian solutions of the unreduced system, which is derived from its associated bilinearized form with nonzero backgrounds. Sec.4 details the classical and nonlocal reduction techniques applied to the Casoratian solutions. The dynamics of solitonic, periodic and rational solutions are illustrated for the focusing sd-mKdV equation and the nonlocal complex defocusing sd-mKdV equation with space-time reversed in Sec.5. The last section is devoted to conclusions.


\section{The unreduced system  and its reductions}

The unreduced system of the coupled sd-mKdV equations  reads
\begin{subequations}\label{sec2-sd-mKdV-coupled1}
\begin{eqnarray}
    &&\partial_tQ_{n}=(1-Q_nR_n)(Q_{n+1}-Q_{n-1}),\\
    &&\partial_tR_{n}=(1-Q_nR_n)(R_{n+1}-R_{n-1}),
\end{eqnarray}
\end{subequations}
which can be derived from a Lax pair \citep{AblowitzL-JMP1976}, i.e., the compatibility condition of the Ablowitz-Ladik (AL) spectral problem
\begin{eqnarray}\label{sec2-spectral}
        \Theta_{n+1}=\mathcal M_n\Theta_n, &&\mathcal M_n=\begin{bmatrix}
            z&Q_n\\
            R_n&z^{-1}
        \end{bmatrix}, ~~~\Theta_n=\begin{bmatrix}
            \theta_{1,n}\\
            \theta_{2,n}
        \end{bmatrix},
    \end{eqnarray}
and the following time evolution equation
\begin{eqnarray}
\label{sec2-time-evo}
        \partial_t\Theta_{n}=\mathcal N_n\Theta_n,&&\mathcal N_n=\frac12\begin{bmatrix}
            z^2-z^{-2}-2Q_nR_{n-1}&2Q_nz+2Q_{n-1}z^{-1}\\
            2R_{n-1}z+2R_nz^{-1}&z^{-2}-z^2-2R_nQ_{n-1}
        \end{bmatrix}.
    \end{eqnarray}
It admits the following reductions
    \begin{subequations}\label{sec2-sd-mKdV-eqs1}
    \begin{eqnarray}
        \partial_tQ_{n}=(1- \delta Q_nQ_n)(Q_{n+1}-Q_{n-1}),&& R_n= \delta Q_n,\label{sec2-sd-mKdV-equ1}\\
        \partial_tQ_{n}=(1- \delta Q_nQ^*_n)(Q_{n+1}-Q_{n-1}),&& R_n= \delta Q_n^*,\label{sec2-sd-mKdV-equ2}\\
        \partial_tQ_{n}=(1-\delta Q_nQ_{-n}(-t))(Q_{n+1}-Q_{n-1}), &&R_n=\delta Q_{-n}(-t),\label{sec2-sd-mKdV-equ3}\\
        \partial_tQ_{n}=(1-\delta Q_nQ^*_{-n}(-t))(Q_{n+1}-Q_{n-1}), &&R_n=\delta Q^*_{-n}(-t).\label{sec2-sd-mKdV-equ4}
    \end{eqnarray}
    \end{subequations}
Here, complex conjugate is denoted by $*$ and the function $Q$ with  reverse-space-time is denoted by $Q_{-n}(-t)=Q(-n,-t)$. Eqs. \eqref{sec2-sd-mKdV-equ1} and \eqref{sec2-sd-mKdV-equ2}, which are obtained with classical reductions, are classical real and complex sd-mKdV equations, respectively.  Eqs. \eqref{sec2-sd-mKdV-equ3} and \eqref{sec2-sd-mKdV-equ4}, which are obtained with nonlocal reduction techniques, are nonlocal real and complex sd-mKdV equations  with  reverse-space-time, respectively.

\section{Bilinearization and Casoratian solutions}
In this section, the construction of the Casoratian solutions to the unreduced system \eqref{sec2-sd-mKdV-coupled1} is shown.

Using the following variable transformation
\begin{eqnarray}\label{sec3-tra1}
    Q_n=G_n/F_n,&&R_n=H_n/F_n,
\end{eqnarray}
the unreduced system \eqref{sec2-sd-mKdV-coupled1}  is transformed to a bilinear system
\begin{subequations}\label{sec3-bilinear1}
    \begin{eqnarray}
        &&(1-q_nr_n)F_{n+1}F_{n-1}+G_nH_n-F_n^2=0,\label{sec3-bilinear-equ1}\\
        &&D_tG_n\cdot F_n=(1-q_nr_n)(G_{n+1}F_{n-1}-G_{n-1}F_{n+1}),\label{sec3-bilinear-equ2}\\
        &&D_tH_n\cdot F_n=(1-q_nr_n)(F_{n+1}H_{n-1}-F_{n-1}H_{n+1}),\label{sec3-bilinear-equ3}
    \end{eqnarray}
    \end{subequations}
where  $(q_n,r_n)$ is a simple solution pair of the unreduced system  \eqref{sec2-sd-mKdV-coupled1} and served as nonzero backgrounds.  The so-called Hirota bilinear operator \citep{Hirota-Book2004} is denoted by $D_t$ and defined as 
    \begin{eqnarray}
        D_t^m G\cdot F=(\partial_t-\partial_{t'})^mG(t)F(t')|_{t'=t}.
    \end{eqnarray}

Note that when $q_n=r_n=0$, this bilinearized system degenerates to that with zero background, which has been shown in \cite{FengZS-IJMPB2020}.

Following \cite{DengCCZ-arXiv2024}, the quasi double Casoratians are constructed as follows:
\begin{subequations}\label{sec3-transform2}
    \begin{eqnarray}
        F(A,\Phi_n,\Psi_n)&=&\alpha_n^{(m+1)/2}|\Phi_{n+1},A^2\Phi_{n+1},...,A^{2m}\Phi_{n+1};\Psi_n,A^2\Psi_n,...,A^{2m}\Psi_n|,\\
        G(A,\Phi_n,\Psi_n)&=&\alpha_n^{(m+1)/2}|\Phi_n,A\Phi_{n+1},A^3\Phi_{n+1},...,A^{2m+1}\Phi_{n+1};A\Psi_n,A^3\Psi_n,...,A^{2m-1}\Psi_n|,\\
        H(A,\Phi_n,\Psi_n)&=&\alpha_n^{(m+1)/2}|A\Phi_{n+1},A^3\Phi_{n+1},...,A^{2m-1}\Phi_{n+1};\Psi_{n+1},A\Psi_n,A^3\Psi_n,...,A^{2m+1}\Psi_n|.
    \end{eqnarray}
\end{subequations}
Here, $\alpha_n=1-q_nr_n$,  the component entries $\Phi_n$ and $\Psi_n$ are  (2m+2)-th order column vectors and given by matrix equations   
    \begin{subequations}\label{sec3-phipsi}
        \begin{align}
            \begin{bmatrix}\Phi_{n+1}\\
                \Psi_{n+1}\end{bmatrix}=M_n\begin{bmatrix}\Phi_{n}\\
                \Psi_{n}\end{bmatrix},\label{sec3-phipsi-spectral} &~~~M_n=\alpha_n^{-1/2}\begin{bmatrix}A&q_nI_{2m+2}\\r_nI_{2m+2}&A^{-1}
                \end{bmatrix},\\
            2\partial_t\begin{bmatrix}\Phi_{n}\\
                \Psi_{n}\end{bmatrix}=N_n\begin{bmatrix}\Phi_{n}\\
                \Psi_{n}\end{bmatrix}, \label{sec3-phipsi-time} &~~~N_n=\begin{bmatrix}{A^2-A^{-2}+(r_nq_{n-1}-q_nr_{n-1})I_{2m+2}}&2Aq_n+2A^{-1}q_{n-1}\\
                    2Ar_{n-1}+2A^{-1}r_{n}&A^{-2}-A^2-(r_nq_{n-1}-q_nr_{n-1})I_{2m+2}
                \end{bmatrix},
        \end{align}
\end{subequations}
    in which $A\in \mathbb{C}_{(2m+2)\times (2m+2)}$, $|A|\neq0$ and $I_{\mu}$ is a $\mu$-th order identity matrix.

    Utilizing these Casoratians, the solutions to the  bilinearized form \eqref{sec3-bilinear1} can be derived, which yields the following theorem.

    \begin{theorem}\label{sec2-theorem1}
        The bilinear system \eqref{sec3-bilinear1} has double Casoratian solutions
            \begin{eqnarray}\label{sec3-transform2}
                F_n=F(A,\Phi_n,\Psi_n), &
                G_n=G(A,\Phi_n,\Psi_n), &
                H_n=H(A,\Phi_n,\Psi_n),
            \end{eqnarray}
    where their vector entries $\Phi_n$ and $\Psi_n$ satisfy matrix equations \eqref{sec3-phipsi} with the given matrix $A$. These solutions also give the Casoratian solutions $Q_n$ and $R_n$ to the unreduced system \eqref{sec2-sd-mKdV-coupled1} through variable transformation \eqref{sec3-tra1}. Matrix $A$ together with its associated vector entries and its similar matrices together with their associated vector entries lead to the same solutions $Q_n$ and $R_n$.
    \end{theorem}
    A brief proof can be find in Appendix \ref{app1-sec1}.

\section{Reductions}\label{sec4}
In this section, the classical and nonlocal reductions to the Casoratian solutions given by theorem \ref{sec2-theorem1} are presented. The constraints between the matrices $A=e^B$ and $T$ induced by associated reductions are discussed, followed by a detail construction of the component vectors $\Phi_n$ and $\Psi_n$.

\subsection{Reductions of the double Casoratians}
In what follows, the detail  reduction procedures to equation \eqref{sec2-sd-mKdV-equ1} are presented. 

Equation \eqref{sec2-sd-mKdV-equ1} is reduced from the unreduced system \eqref{sec2-sd-mKdV-coupled1} by requiring $R_n=\delta Q_n$, hence the nonzero backgrounds are assumed to satisfy $r_n=\delta q_n$.
By imposing constraint
\begin{eqnarray}\label{sec3-phipsi-reduction1-case1}
    \Psi_n=T\Phi_{n},
\end{eqnarray}
and assuming
\begin{eqnarray}\label{sec3-at-relation-case1}
      A^{-1}T=TA, & TT=\delta I_{2m+2},
\end{eqnarray}
one can verify that the matrix system reduces to the following system
\begin{subequations}\label{sec3-phipsi-reduction2-case1}
    \begin{align}\label{sec3-phipsi-reduction2-spectral-case1}
        \Phi_{n+1}=\alpha_n^{-1/2}A\Phi_n+\alpha_n^{-1/2}q_nT\Phi_{n},\\
        2\partial_t\Phi_{n}=(A^2-A^{-2})\Phi_n+2(Aq_n+A^{-1}q_{n-1})T\Phi_{n}
    \end{align}
\end{subequations}
with $\alpha_n=1-\delta q_n^2$. If introducing $B$ such that $A=e^B$, the assumption \eqref{sec3-at-relation-case1} becomes
\begin{eqnarray}\label{sec3-at-relation2-case1}
  -BT=TB, & TT=\delta I_{2m+2}.
\end{eqnarray}

Using matrix equation \eqref{sec3-phipsi-spectral} and constraint \eqref{sec3-phipsi-reduction1-case1}, the double Casoratian $F_n$ can be written as 
\begin{eqnarray}
    F_n=|A\Phi_n,...,A^{2m+1}\Phi_n;\Psi_n,...,A^{2m}\Psi_n|=|A\Phi_n,...,A^{2m+1}\Phi_n;T\Phi_n,...,A^{2m}T\Phi_n|.
 \end{eqnarray}
 With constraints \eqref{sec3-phipsi-reduction1-case1} and  \eqref{sec3-at-relation-case1}, one can see that
 \begin{eqnarray}
    F_n=|A^{2m+1}T|^{-1}|A^{2m}T\Phi_n,...,T\Phi_n;\delta A^{2m+1}\Phi_n,...,\delta A\Phi_n|=(-\delta)^{m+1}|A^{2m+1}T|^{-1}F_n.
 \end{eqnarray}

 Similarly, we can also derive that $H_n=- (-\delta)^{m}|A^{2m+1}T|^{-1}G_n$. They give the relation between $Q_n$ and $R_n$ by
 \begin{eqnarray}
     R_n=\frac{H_n}{F_n}=\frac{-(-\delta)^m|A^{2m+1}T|^{-1}G_n}{(-\delta)^{m+1}|A^{2m+1}T|^{-1}F_n}=\delta G_n/F_n=\delta Q_n,
 \end{eqnarray}
 which implies that $Q_n$ is a solution of the  sd-mKdV equation \eqref{sec2-sd-mKdV-equ1}.

This reduction procedure can also be applied to equations \eqref{sec2-sd-mKdV-equ2}, \eqref{sec2-sd-mKdV-equ3} and \eqref{sec2-sd-mKdV-equ4}. We skip the details and list main results in Table \ref{table-theo2}. Note that in this table, vector $\Phi_n$ for \eqref{sec2-sd-mKdV-equ2}, \eqref{sec2-sd-mKdV-equ3} and \eqref{sec2-sd-mKdV-equ4} is determined by

\begin{subequations}\label{sec3-phipsi-reduction2-case2}
    \begin{align}\label{sec3-phipsi-reduction2-spectral-case2}
        \Phi_{n+1}=\alpha_n^{-1/2}A\Phi_n+\alpha_n^{-1/2}q_nT\Phi^*_n,\\
        2\partial_t\Phi_n=(A^2-A^{-2}-\delta q_nq^*_{n-1}+\delta q_{n-1}q^*_{n})\Phi_n+2(Aq_n+A^{-1}q_{n-1})T\Phi^*_n
    \end{align}
\end{subequations}
with $\alpha_n=1-\delta |q_n|^2$,
\begin{subequations}\label{sec3-phipsi-reduction2-case3}
    \begin{align}\label{sec3-phipsi-reduction2-spectral-case3}
        \Phi_{n+1}=\alpha_n^{-1/2}A\Phi_n+\alpha_n^{-1/2}q_nT\Phi_{1-n}(-t),\\
        2\partial_t\Phi_{n}=(A^2-A^{-2}-\delta q_nq_{1-n}(-t)+\delta q_{n-1}q_{-n}(-t))\Phi_n+2(Aq_n+A^{-1}q_{n-1})T\Phi_{1-n}(-t)
    \end{align}
\end{subequations}
with $\alpha_n=1-\delta q_nq_{-n}(-t)$ and
\begin{subequations}\label{sec3-phipsi-reduction2-case4}
    \begin{align}\label{sec3-phipsi-reduction2-spectral-case4}
        \Phi_{n+1}=\alpha_n^{-1/2}A\Phi_n+\alpha_n^{-1/2}q_nT\Phi^*_{1-n}(-t),\\
        2\partial_t\Phi_{n}=(A^2-A^{-2}-\delta q_nq_{1-n}^*(-t)+\delta q_{n-1}q_{-n}^*(-t))\Phi_n+2(Aq_n+A^{-1}q_{n-1})T\Phi_{1-n}^*(-t)
    \end{align}
\end{subequations}
with $\alpha_n=1-\delta q_nq_{-n}^*(-t)$, respectively.

The above results are summarized  as the following theorem.

\begin{theorem}\label{theorem-red1}
Solutions of the four sd-mKdV equations in \eqref{sec2-sd-mKdV-eqs1} are given by variable transformation
\begin{eqnarray}
    Q_n=\frac{G_n}{F_n},
\end{eqnarray}
in which $G_n$ and $F_n$ are double Casoratians defined in formulae \eqref{sec3-transform2} with their vector entries $\Phi_n$ and $\Psi_n$ given by Table \ref{table-theo2}.

\end{theorem}

\begin{table}[htbp]
    \centering
    \captionsetup{font={small}}
    \caption{Reduction constraints for equations \eqref{sec2-sd-mKdV-eqs1}\label{table-theo2}}
    \begin{tabular}{|c|c|c|c|c|}
    \hline
                                equation  &            $(q_n,r_n)$           &  constraints &  $F_n,~G_n,~H_n$              &  $\Phi_n$                                  \\
    \hline
    \multirow{3}{*}{\eqref{sec2-sd-mKdV-equ1}} &\multirow{3}{*}{$r_n=\delta q_n$}  &
    $\Psi_n=T\Phi_{n}$ &  $F_n=(-\delta)^{m+1}|A^{2m+1}T|^{-1}F_n$ & \multirow{3}{*}{\eqref{sec3-phipsi-reduction2-case1}}\\
&& $ A^{-1}T=TA,~~ TT=\delta I_{2m+2}$ & $H_n=-(-\delta)^{m}|A^{2m+1}T|^{-1}G_n$ &  \\
&& $ -BT=TB,~~ TT=\delta I_{2m+2}$ & &\\
\hline
\multirow{3}{*}{\eqref{sec2-sd-mKdV-equ2}} &\multirow{3}{*}{$r_n=\delta q^*_n$}  &
$\Psi_n=T\Phi^*_{n}$ &  $F_n=(-\delta)^{m+1}|A^{2m+1}T|^{-1}F^*_n$ & \multirow{3}{*}{\eqref{sec3-phipsi-reduction2-case2}}\\
&& $ A^{-1}T=TA^*,~~ TT^*=\delta I_{2m+2}$ & $H_n=-(-\delta)^{m}|A^{2m+1}T|^{-1}G^*_n$ &  \\
&& $ -BT=TB^*,~~ TT^*=\delta I_{2m+2}$ & &\\
\hline
\multirow{3}{*}{\eqref{sec2-sd-mKdV-equ3}} &\multirow{3}{*}{$r_n=\delta q_{-n}(-t)$}  &
$\Psi_n=T\Phi_{1-n}(-t)$ &  $F_n=\delta^{m+1}|T|^{-1}F_{-n}(-t)$ & \multirow{3}{*}{\eqref{sec3-phipsi-reduction2-case3}}\\
&& $ AT=TA,~~ TT=-\delta I_{2m+2}$ & $H_n=\delta^{m}|T|^{-1}G_{-n}(-t)$ &  \\
&& $ BT=TB,~~ TT=-\delta I_{2m+2}$ & &\\
\hline
\multirow{3}{*}{\eqref{sec2-sd-mKdV-equ4}} &\multirow{3}{*}{$r_n=\delta q^*_{-n}(-t)$}  &
$\Psi_n=T\Phi^*_{1-n}(-t)$ &  $F_n=\delta^{m+1}|T|^{-1}F^*_{-n}(-t)$ & \multirow{3}{*}{\eqref{sec3-phipsi-reduction2-case4}}\\
&& $ AT=TA^*,~~ TT^*=-\delta I_{2m+2}$ & $H_n=\delta^{m}|T|^{-1}G^*_{-n}(-t)$ &  \\
&& $ BT=TB^*,~~ TT^*=-\delta I_{2m+2}$ & &\\
\hline
    \end{tabular}
    \end{table}

\begin{rmk}
    By defining $\hat T=\gamma A^{\mu}T$, the  constraint \eqref{sec3-phipsi-reduction1-case1}  becomes $\gamma A^{\mu}\Psi_n=\hat T\Phi_n$, the assumption \eqref{sec3-at-relation-case1} yields ${A}^{-1}{\hat T}={\hat T}{A}$ and ${\hat T}{\hat T}=\gamma^{2}\delta I_{2m+2}$ and the Casoratians change accordingly, where $\mu$ is an arbitrary integer and $\gamma$ is an arbitrary nonzero complex number.  Similar discussions can be applied to other equations in \eqref{sec2-sd-mKdV-eqs1}, which allow us to obtain Casoratians presented in \cite{FengZS-IJMPB2020} when requiring $q_n=r_n=0$ and choosing $\gamma$ and $\mu$ properly.  
\end{rmk}

\subsection{Matrices $B$ and $T$}
This subsection discusses the explicit forms of $B$ and $T$ in theorem \ref{theorem-red1}.

Assuming matrices $B$ and $T$ admit the following specific form 
\begin{eqnarray}\label{sec4-bt-sol}
    B=\begin{bmatrix}
        K_1&\mathbf{0}\\
        \mathbf{0}&K_4
    \end{bmatrix}, && T=\begin{bmatrix}
        T_1&T_2\\
        T_3&T_4
    \end{bmatrix},
\end{eqnarray}
where $T_\mu$ and $K_\mu$ are $(m+1)\times (m+1)$ matrices,  some solutions to the constraints given by Table \ref{table-theo2}  are  presented in Table \ref{table-at}.

\begin{table}[htbp]
    \centering
    \captionsetup{font={small}}
    \caption{Matrices $B$ and~$T$ in constraints for equations \eqref{sec2-sd-mKdV-eqs1}}\label{table-at}
    \begin{tabular}{|c|c|c|c|c|}
    \hline
                           equation       &            $\delta$           &  $T$              &  $B$                                  \\
    \hline
    {\eqref{sec2-sd-mKdV-equ1}} & $\delta$  &
    $T_{1}=T_{4}=\mathbf{0}_{m+1},T_{2}=\delta T_{3}=\mathbf{I}_{m+1}$  &
    {$K_{1}=\mathbf{K}_{m+1}, K_{4}=-\mathbf{K}_{m+1}$}    \\
        \cline{1-4}

    {\eqref{sec2-sd-mKdV-equ2}} & $\delta$  &
    $T_{1}=T_{4}=\mathbf{0}_{m+1},T_{2}=\delta T_{3}=\mathbf{I}_{m+1}$  &
    {$K_{1}=\mathbf{K}_{m+1}, K_{4}=-\mathbf{K}^*_{m+1}$}    \\
        \cline{1-4}
        \multirow{2}{*}{\eqref{sec2-sd-mKdV-equ3}} & $1$ &
    $T_{1}=-T_{4}=i\mathbf{I}_{m+1},T_{2}=T_{3}=\mathbf{0}_{m+1}$ &
    \multirow{2}{*}{$K_{1}=\mathbf{K}_{m+1}, K_{4}=\mathbf{H}_{m+1}$} \\
        \cline{2-3}
                                     &  $-1$ & $T_{1}=-T_{4}=\mathbf{I}_{m+1},T_{2}=T_{3}=\mathbf{0}_{m+1}$  &  \\
        \cline{1-4}
        \hline
    {\eqref{sec2-sd-mKdV-equ4}} & $\delta$ &
    $T_{1}=T_{4}=\mathbf{0}_{m+1},T_{2}=-\delta T_{3}=\mathbf{I}_{m+1}$ &
    {$K_{1}=\mathbf{K}_{m+1}, K_{4}=\mathbf{K}^*_{m+1}$} \\
        \cline{1-4}
    \hline
    \end{tabular}
    \end{table}

Additionally, the constraint for equation \eqref{sec2-sd-mKdV-equ4} with $\delta=1$ admits a real solution of form \eqref{sec4-bt-sol} with
\begin{subequations}\label{sec4-bt-sol-spe1}
    \begin{eqnarray}
        &&K_1=\mathbf{K}_{m+1},~~K_4=\mathbf{H}_{m+1}, ~~~\mathbf{K}_{m+1},\mathbf{H}_{m+1}\in \mathbb{R}_{(m+1)\times(m+1)},\\
        &&T_1=\pm T_4=I_{m+1},~~ T_2=T_3=\mathbf{0}_{m+1}.
    \end{eqnarray}
\end{subequations}

Using theorem \ref{sec2-theorem1} and $A=e^B$,
it is easy to see that matrix $B$ and its similar matrices can lead to the same solution $Q_n$ of the reduced equations \eqref{sec2-sd-mKdV-eqs1}, which allow us to further restrict the discussions of matrix $B$ to its canonical form, namely matrix $\mathbf K_{m+1}$ is assumed to be
    \begin{eqnarray}
        \mathbf{K}_{m+1}=Diag(J_{\mu_1}(k_1),J_{\mu_2}(k_2),...,J_{\mu_s}(k_s)),
    \end{eqnarray}
where $J_{\mu}(k)$ is Jordan block defined by
\begin{eqnarray}\label{Jordan}
        J_{h}(k)=\begin{bmatrix}
        k & 0 & 0 & \ldots & 0 & 0\\
        1 & k & 0 & \ldots & 0 & 0\\
        \ldots & \ldots & \ldots & \ldots & \ldots & \ldots\\
        0 & 0 & 0 & \ldots & 1 & k
        \end{bmatrix}_{\mu\times \mu},
\end{eqnarray}
with $\sum_{j=1}^s\mu_j=m+1$. 

There are two extreme cases of the canonical forms gained much attention. The first case assumes that all $\mu_j=1$, which results to following diagonal matrix
\begin{eqnarray}
    \mathbf{K}_{m+1}=\textrm{Diag}(k_1,k_2,...,k_{m+1}),
\end{eqnarray}
while the second case assumes that $\mu_1=m+1$, resulting to the following Jordan matrix
\begin{eqnarray}
    \mathbf{K}_{m+1}=J_{m+1}(k_1).
\end{eqnarray}

\subsection{Vectors $\Phi_n$ and $\Psi_n$}
To construct component vectors $\Phi_n$ and $\Psi_n$, the following system 
\begin{subequations}\label{sec4-phipsi}
    \begin{align}
        \begin{bmatrix}\phi_{n+1}\\
            \psi_{n+1}\end{bmatrix}=\alpha_n^{-1/2}\begin{bmatrix}e^k&q_n\\r_n&e^{-k}
            \end{bmatrix}\begin{bmatrix}\phi_{n}\\
            \psi_{n}\end{bmatrix},\label{sec4-phipsi-spectral}\\
        2\begin{bmatrix}\phi_{n}\\
            \psi_{n}\end{bmatrix}_t=\begin{bmatrix}e^{2k}-e^{-2k}+(r_nq_{n-1}-q_nr_{n-1})&2e^kq_n+2e^{-k}q_{n-1}\\
                2e^kr_{n-1}+2e^{-k}r_{n}&e^{-2k}-e^{2k} -(r_nq_{n-1}-q_nr_{n-1})
            \end{bmatrix}\begin{bmatrix}\phi_{n}\\
            \psi_{n}\end{bmatrix},\label{sec4-phipsi-time}
    \end{align}
    \end{subequations}
which can be seen as a scalar version of the matrix system \eqref{sec3-phipsi}, is considered.

Assume that the nonzero backgrounds are given by
\begin{eqnarray}
    q_n=a_0, && r_n=b_0,
\end{eqnarray}

then the system \eqref{sec4-phipsi} admits the following solution pair
\begin{subequations}\label{phipsi-sol-pairs1}
    \begin{eqnarray}
        &&\phi_n(k,c,d)=ce^{\lambda n+\eta t}+de^{-(\lambda n+\eta t)},\\
        &&\psi_n(k,c,d)=-c\xi(k)e^{\lambda n+\eta t}+d\xi(-k) e^{-(\lambda n+\eta t)},
    \end{eqnarray}
    with
    \begin{eqnarray}\label{phipsi-sol-pairs2-1}
        &&e^\lambda=e^{\lambda(k)}=\frac{e^k+e^{-k}+\sqrt{(e^k-e^{-k})^2+4a_0b_0}}{2\sqrt{1-a_0b_0}},\\ \label{phipsi-sol-pairs2-2}
        &&\eta=\eta(k)=\frac 12(e^k+e^{-k})\sqrt{(e^k-e^{-k})^2+4a_0b_0},\\ \label{phipsi-sol-pairs2-3}
        &&\xi=\xi(k)=\frac{e^k-e^{-k}-\sqrt{(e^k-e^{-k})^2+4a_0b_0}}{2a_0}. 
    \end{eqnarray}
    \end{subequations}
    Here $c$ and $d$ are constants (or functions of $k$). To simplify discussions, assumptions $b_0= \delta a_0$ and $a_0  \in \mathbb R$ are taken.
    
    In what follows, the construction of vectors $\Phi_n$ and $\Psi_n$ using the above solution pair $\phi_n(k,c,d)$ and $\psi_n(k,c,d)$ is shown.

\subsubsection{The unreduced case}\label{sec4-cas1}
Defining 
\begin{subequations}\label{hat-check-phipsi1}
\begin{eqnarray}
    &&\Phi_n=(\phi_n(k_1,c_1,d_1),\phi_n(k_2,c_2,d_2),...,\phi_n(k_{2m+2},c_{2m+2},d_{2m+2}))^T,\\
    &&\Psi_n=(\psi_n(k_1,c_1,d_1),\psi_n(k_2,c_2,d_2),...,\psi_n(k_{2m+2},c_{2m+2},d_{2m+2}))^T,
\end{eqnarray}
\end{subequations}
the resulting double Casoratians provide solutions of the unreduced system \eqref{sec2-sd-mKdV-coupled1} via the transform \eqref{sec3-tra1}  with regards to matrix $B=\textrm{Diag}(k_1,k_2,...,k_{2m+2})$.

While defining
\begin{subequations}\label{hat-check-phipsi2}
\begin{eqnarray}
    &&\Phi_n=(\phi_n(k_1,c_1,d_1),\frac{\partial_{k_1}}{1!}\phi_n(k_1,c_1,d_1),...,\frac{\partial^{2m+1}_{k_1}}{(2m+1)!}\phi_n(k_1,c_1,d_1))^T,\\
    &&\Psi_n=(\psi_n(k_1,c_1,d_1),\frac{\partial_{k_1}}{1!}\psi_n(k_1,c_1,d_1),...,\frac{\partial^{2m+1}_{k_1}}{(2m+1)!}\psi_n(k_1,c_1,d_1))^T,
\end{eqnarray}
\end{subequations}
the resulting double Casoratians give solutions corresponding to matrix $B=J_{2m+2}(k_1)$.

\subsubsection{The reduced cases}\label{sec4-cas2}
To discuss the reduced equations \eqref{sec2-sd-mKdV-eqs1}, we define
\begin{subequations}\label{sec4-phipsi-relation}
\begin{eqnarray}
&&\Phi_n=(\Phi_n^+,\Phi_n^-)^T,~~~ \Phi_n^\pm=(\Phi_{1,n}^\pm,...,\Phi_{m+1,n}^\pm)^T,\\
&&\Psi_n=(\Psi_n^+,\Psi_n^-)^T,~~~ \Psi_n^\pm=(\Psi_{1,n}^\pm,...,\Psi_{m+1,n}^\pm)^T,
\end{eqnarray}
\end{subequations}
with $\Phi_{j,n}^\pm$ and $\Psi_{j,n}^\pm$ being scalar functions.

For equations \eqref{sec2-sd-mKdV-equ1},  \eqref{sec2-sd-mKdV-equ2} and \eqref{sec2-sd-mKdV-equ4}, the matrix $T$ in constraints is block skew diagonal, which allows us to  express $\Phi_n$ and  $\Psi_n$ with $\Phi_n^+$ and  $\Psi_n^+$. The detail expressions are listed in Table \ref{table-phipsi}. In this scenario, vectors $\Phi_n^+$ and  $\Psi_n^+$ are given by

\begin{eqnarray}
    \Phi_{j,n}^+=\phi_n(k_j^+,c_j^+,d_j^+), & \Psi_{j,n}^+=\psi_n(k_j^+,c_j^+,d_j^+),& j=1,2,...,m+1,
\end{eqnarray}
when $\mathbf K_{m+1}=\textrm{Diag}(k_1^+,k_2^+,...,k_{m+1}^+)$ is a diagonal matrix,
or are given by
\begin{eqnarray}
    \Phi_{j,n}^+=\frac{\partial^{j-1}_{k_1^+}\phi(k_1^+,c_1^+,d_1^+)}{(j-1)!}, &   \Psi_{j,n}^+=\frac{\partial^{j-1}_{k_1^+}\psi(k_1^+,c_1^+,d_1^+)}{(j-1)!}, & j=1,2,...,m+1,
\end{eqnarray}
when $\mathbf K_{m+1}= J_{m+1}( k_1^+)$ is a Jordan matrix.






For equation \eqref{sec2-sd-mKdV-equ3} and the special reduction \eqref{sec4-bt-sol-spe1}, the matrix $T$ in constraints is block diagonal, which enable us to express   $\Psi_n$ in terms of $\Phi_n$ with appropriate restrictions between the parameters $c$ and $d$. The detail expressions and restrictions are listed in Table \ref{table-phipsi2}. In this scenario, vector $\Phi_n$ is given by

\begin{eqnarray}
    \Phi_{j,n}^+=\phi_n(k_j^+,c_j^+,d_j^+), & \Phi_{j,n}^-=\psi_n(h_j^-,c_j^-,d_j^-),& j=1,2,...,m+1,
\end{eqnarray}
when both $\mathbf K_{m+1}=\textrm{Diag}(k_1^+,k_2^+,...,k_{m+1}^+)$ and $\mathbf H_{m+1}=\textrm{Diag}(h_1^-,h_2^-,...,h_{m+1}^-)$ are diagonal,
or is given by
\begin{eqnarray}
    \Phi_{j,n}^+=\frac{\partial^{j-1}_{k_1^+}\phi(k_1^+,c_1^+,d_1^+)}{(j-1)!}, &   \Phi_{j,n}^-=\frac{\partial^{j-1}_{h_1^-}\psi(h_1^-,c_1^-,d_1^-)}{(j-1)!}, & j=1,2,...,m+1,
\end{eqnarray}
when both $\mathbf K_{m+1}= J_{m+1}( k_1^+)$ and $\mathbf K_{m+1}= J_{m+1}( h_1^-)$ are  Jordan matrices.



\begin{table}[htbp]
    \centering
    \captionsetup{font={small}}
    \caption{Vectors $\Phi_n$ and $\Psi_n$ when matrix $T$ is block skew diagonal}\label{table-phipsi}
    \begin{tabular}{|c|c|c|c|c|}
    \hline
                           equation       &            $\Phi_n$           &  $\Psi_n$                                               \\
    \hline
    \eqref{sec2-sd-mKdV-equ1}    &            $(\Phi^+_n,\Psi_n^+)^T$           &  $(\Psi^+_n,\delta\Phi_n^+)^T$                                                \\
    \hline
    \eqref{sec2-sd-mKdV-equ2}    &            $(\Phi^+_n,\Psi_n^{+*})^T$           &  $(\Psi^+_n,\delta\Phi_n^{+*})^T$                                                \\
    \hline
    \eqref{sec2-sd-mKdV-equ4}    &            $(\Phi^+_n,\Psi_{1-n}^{+*}(-t))^T$           &  $(\Psi^+_n,-\delta\Phi_{1-n}^{+*}(-t))^T$                                              \\
    \hline
    \end{tabular}
\end{table}

\begin{table}[htbp]
    \centering
    \captionsetup{font={small}}
    \caption{Vectors $\Phi_n$ and $\Psi_n$ when matrix $T$ is block diagonal}\label{table-phipsi2}
    \begin{tabular}{|c|c|c|c|c|}
    \hline
                           equation       & $\delta$ &            $\Phi_n$           &  $\Psi_n$      & restriction between $c$ and $d$                                         \\
    \hline
    \multirow{4}{*}{\eqref{sec2-sd-mKdV-equ3}}   & \multirow{2}{*}{$1$} &            \multirow{2}{*}{$(\Phi^+_n,\Phi_n^-)^T$}           &  \multirow{2}{*}{$(i\Phi^+_{1-n}(-t),-i\Phi^-_{1-n}(-t))^T$}   &  $d_j^+=i\xi(k_j^+)e^{\lambda(k_j^+)} c^+_j$                                           \\
    & & & & $d_j^-=-i\xi(h_j^-)e^{\lambda(h_j^-)} c^-_j$\\
    \cline{2-5}
    & \multirow{2}{*}{$-1$} &            \multirow{2}{*}{$(\Phi^+_n,\Phi_n^-)^T$}           &  \multirow{2}{*}{$(\Phi^+_{1-n}(-t),-\Phi^-_{1-n}(-t))^T$}   &  $d_j^+=-\xi(k_j^+)e^{\lambda(k_j^+)} c^+_j$                                           \\
    &&&&$d_j^-=\xi(h_j^-)e^{\lambda(h_j^-)} c^-_j$\\
    \hline
    \multirow{2}{*}{\eqref{sec4-bt-sol-spe1} for \eqref{sec2-sd-mKdV-equ4}} & \multirow{2}{*}{$1$}   &            \multirow{2}{*}{$(\Phi^+_n,\Phi_n^-)^T$}           &  \multirow{2}{*}{$(\Phi^{+*}_{1-n}(-t),\pm\Phi^{-*}_{1-n}(-t))^T$}   &  $d_j^+= -\xi(k_j^+)e^{\lambda(k_j^+)} c^+_j$                                           \\
    & & & & $d_j^-=\mp\xi(h_j^-)e^{\lambda(h_j^-)} c^-_j$\\
    \hline

    \end{tabular}
\end{table}

\section{Dynamics}\label{sec5}
This section devotes to demonstrating dynamics such as solitons, periodic waves and rational solutions. 
\subsection{The focusing sd-mKdV equation}\label{sec51}
\subsubsection{solitonic and periodic solutions}
When $m=0$ and $\mathbf K_1=k_1$,  the first order solution $Q_n=G_n/F_n$ for the focusing sd-mKdV equation \eqref{sec2-sd-mKdV-equ1}  ($\delta=-1$) can be presented by  Casoratians
\begin{subequations}\label{sec5-sol1}    
    \begin{eqnarray}
        &&F_n=-\sqrt{1+a_0^2}\Bigr[A_1e^{2\lambda_1 n+2\eta_1 t}+B_1e^{-2\lambda_1 n-2\eta_1 t}+C_1\Bigr],\\
        &&G_n=\sqrt{1+a_0^2}\Bigr[A_2e^{2\lambda_1 n+2\eta_1 t}+B_2e^{-2\lambda_1 n-2\eta_1 t}+C_2\Bigr],
    \end{eqnarray}
with
    \begin{align}
        \lambda_1=\lambda(k_1),~~ &\eta_1=\eta(k_1),\\
        A_1=c_1^2e^{\lambda_1}(1+\xi(k_1)\xi(k_1)),~~ &A_2=c_1^2e^{\lambda_1}\xi(k_1)(e^{k_1}-e^{-k_1}),\\
        B_1=d_1^2e^{-\lambda_1}(1+\xi(-k_1)\xi(-k_1)),~~ &B_2=d_1^2e^{-\lambda_1}\xi(-k_1)(e^{-k_1}-e^{k_1}), \\
        C_1=2c_1d_1(e^{\lambda_1}+e^{-\lambda_1}),~~&      C_2=c_1d_1(e^{-\lambda_1-k_1}\xi(-k_1)-e^{\lambda_1-k_1}\xi(k_1)-e^{\lambda_1+k_1}\xi(-k_1)+e^{-\lambda_1+k_1}\xi(k_1)).
    \end{align}
\end{subequations}
Here, $\lambda(k), \eta(k)$ and $\xi(k)$ are defined in \eqref{phipsi-sol-pairs2-1}-\eqref{phipsi-sol-pairs2-3} with $b_0=-a_0$.
It is easy to see that the solution dynamics are controlled by $X_1=\lambda_1 n+\eta_1 t$. We are interested in two special cases. 

\begin{figure}[ht]
\centering
\subfigure[ ]{
\begin{minipage}[t]{0.32\linewidth}
\centering
\includegraphics[width=2.0in]{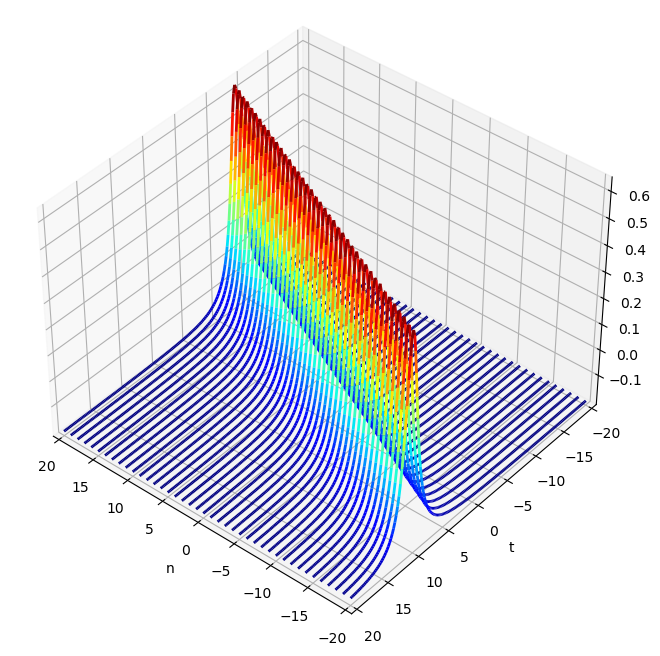}
\end{minipage}\label{fig-classical-soliton-sol1}
}%
\subfigure[ ]{
\begin{minipage}[t]{0.32\linewidth}
\centering
\includegraphics[width=2.0in]{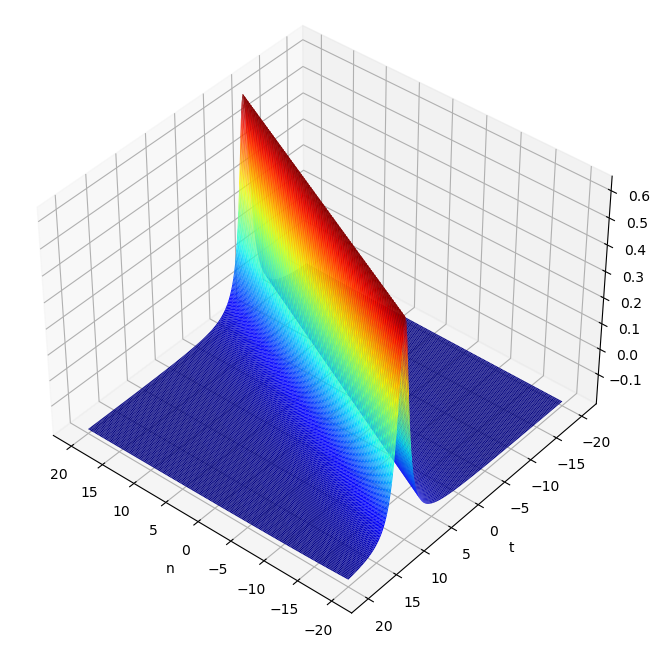}
\end{minipage}\label{fig-classical-soliton-sol2}
}%
\subfigure[ ]{
\begin{minipage}[t]{0.32\linewidth}
\centering
\includegraphics[width=2.0in]{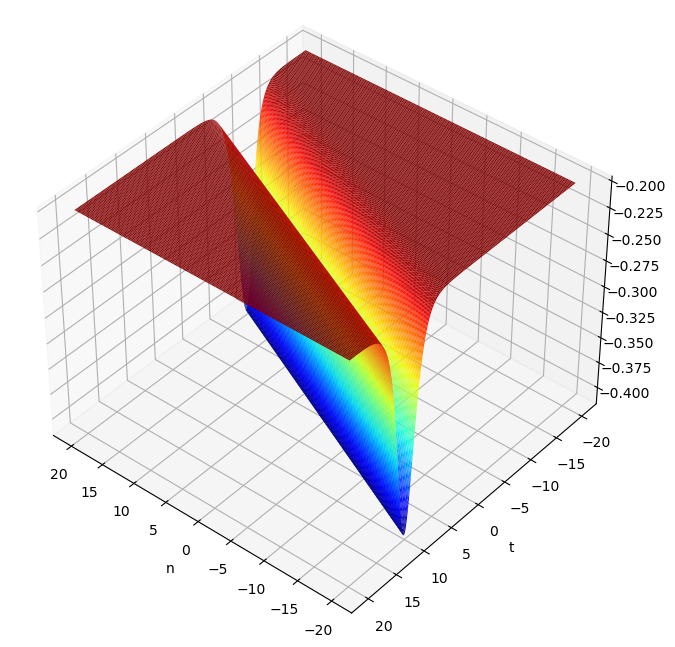}
\end{minipage}\label{fig-classical-soliton-sol3}
}%

\caption{\label{fig-classical-soliton-sols1} Shape and motion of first order solutions ($Q_n$ is presented) for the focusing sd-mKdV equation. (a) and (b) A soliton with $a_0=0.2,~ k_1=0.2,~ c_1=-d_1=1$.    (c) An anti-soliton for $a_0=0.2,~ k_1=0.3,~ c_1=d_1=1$.   
}
\end{figure}

The first case is that 
\begin{eqnarray}\label{sec5-conditions-sol1}
    k_1, c_1, d_1 \in \mathbb{R}, && (e^{k_1}-e^{-k_1})^2>4a_0^2,
\end{eqnarray}
which yields real $\lambda_1$, $\eta_1$ and $Q_n$. For a fixed $t$, we have
\begin{eqnarray}
    \lim_{\lambda_1n\rightarrow\pm\infty}Q_n=a_0,
\end{eqnarray}
which tells it lives on a nonzero background given by $a_0$. 
Further analyzing the possible maximum (minimum), one can obtain soliton (anti-soliton) characterized by its maximum (minimum)  traveling along $X_1=\frac{1}{2}\ln|\frac{d_1}{c_1\xi(k_1)e^{\lambda_1}}|$ with constant traveling speed $\eta_1/\lambda_1$. A soliton is shown  Fig. \ref{fig-classical-soliton-sol1} and \ref{fig-classical-soliton-sol2}\footnote[1]{We should in principle demonstrate all figures as Fig. \ref{fig-classical-soliton-sol1}, but in what follows we show them as Fig. \ref{fig-classical-soliton-sol2} to  visualize the solution dynamics more clearly.}, while an anti-soliton is shown in Fig. \ref{fig-classical-soliton-sol3}.

\captionsetup[figure]{labelfont={bf},name={Fig.},labelsep=period}

\begin{figure}[ht]
\centering
\begin{minipage}[t]{0.32\linewidth}
\centering
\includegraphics[width=2.0in]{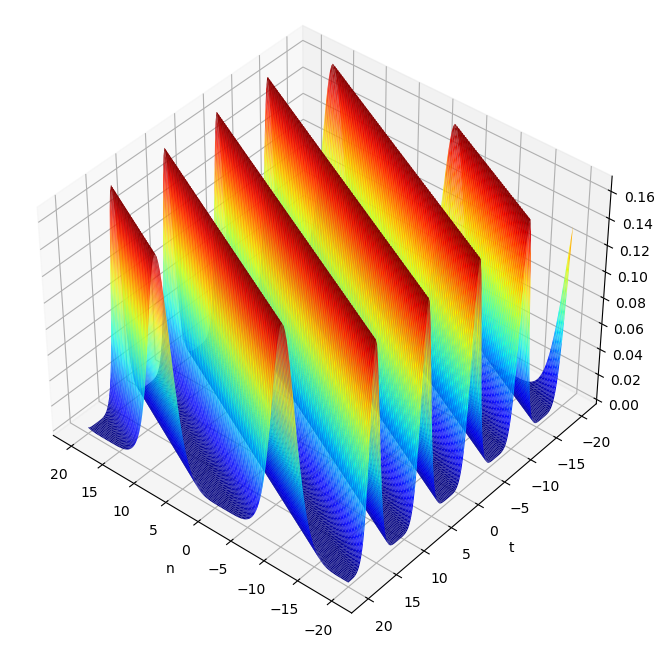}
\end{minipage}%

\caption{\label{fig-classical-wave-sols2} Shape and motion of first order doubly periodic solutions ($|Q_n|^2$ is presented) for the focusing sd-mKdV equation with $a_0=0.2,~ k_1=0.1,~ c_1=d_1=1$.}
\end{figure}

The second case is when
\begin{eqnarray}\label{sec5-conditions-wav1}
     k_1 \in \mathbb{R}, &&(e^{k_1}-e^{-k_1})^2<4a_0^2, 
\end{eqnarray}
or 
\begin{eqnarray}\label{sec5-conditions-wav2}
     k_1 \in i\mathbb{R},
\end{eqnarray}
which yields that both $\lambda_1$ and  $\eta_1$ are purely imaginary. In this scenario, the solution $Q_n$ is doubly periodic with period $T_1=|2\pi /\lambda_1|$  in $n$-direction \footnote[2]{In principle $T$ is a period only when it is a positive integer due the discretized spatial coordinate $n$. But we also call it period for convenient.}  and  $T_2=|2\pi/\eta_1|$ in $t$-direction. An example is shown in Fig. \ref{fig-classical-wave-sols2}.

\captionsetup[figure]{labelfont={bf},name={Fig.},labelsep=period}

\begin{figure}[ht]
\centering
\subfigure[ ]{
\begin{minipage}[t]{0.32\linewidth}
\centering
\includegraphics[width=2.0in]{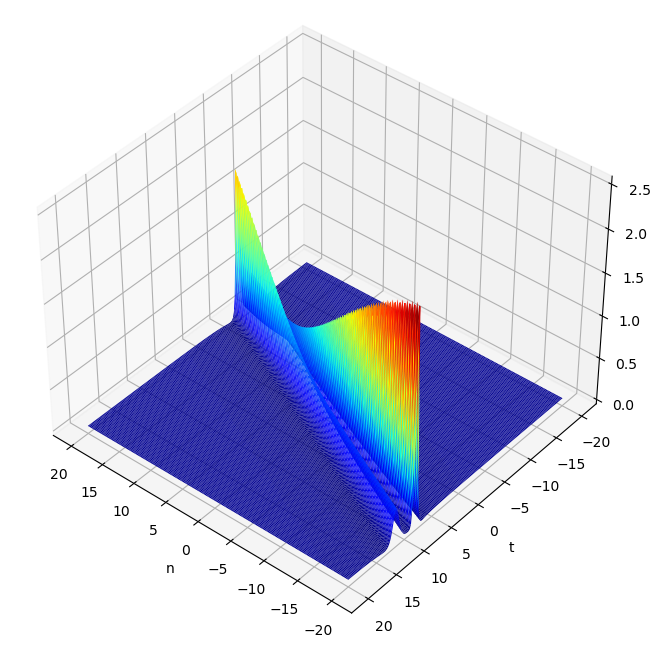}
\end{minipage}%
}%
\subfigure[ ]{
\begin{minipage}[t]{0.32\linewidth}
\centering
\includegraphics[width=2.0in]{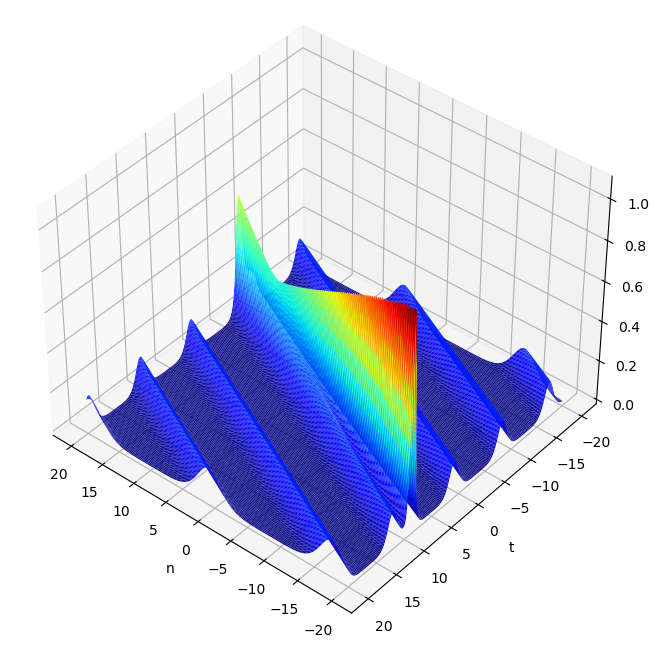}
\end{minipage}%
}%
\subfigure[ ]{
\begin{minipage}[t]{0.32\linewidth}
\centering
\includegraphics[width=2.0in]{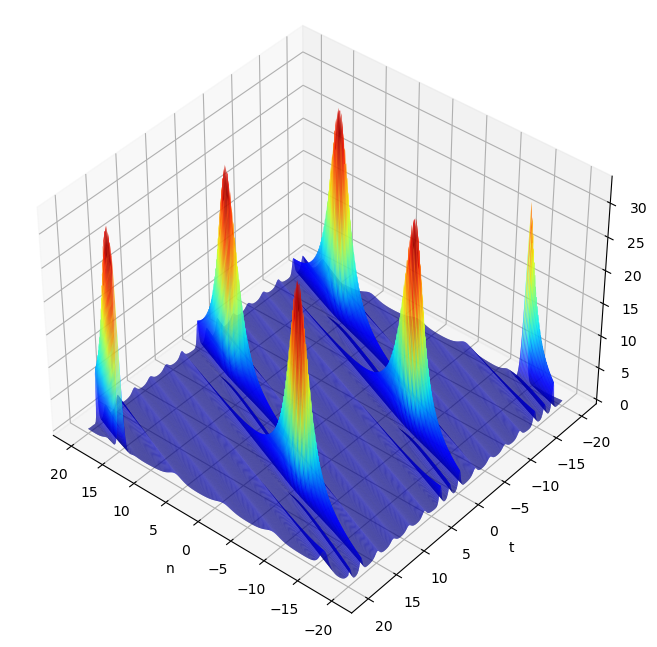}
\end{minipage}%
}%

\caption{\label{fig-classical-diag-sols3}Shape and motion of second order solutions (diagonal matrix, $|Q_n|^2$ is presented) for the focusing sd-mKdV equation.  (a)  Interaction between two soltions with $a_0=0.2,~ k_1=0.2,~ k_2=0.6,~ c_1=-d_1=c_2=d_2=1$. (b) Interaction between soliton and doubly periodic wave with  $a_0=0.2,~ k_1=0.1,~ k_2=0.5,~ c_1=-d_1=c_2=d_2=1$.  (c) Interaction between two doubly periodic waves with $a_0=0.7,~ k_1=0.4,~ k_2=0.5,~ c_1=-d_1=c_2=d_2=1$.   
}
\end{figure}

When $m=1$ and $\mathbf K_2=diag(k_1,k_2)$ is  diagonal,  the second order solution $Q_n=G_n/F_n$ can be investigated by using  the following Casoratians
\begin{eqnarray}\label{sec5-sols3-det1}
    F_n=\alpha_n|\Phi_{n+1},A^2\Phi_{n+1},T\Phi_{n},A^2T\Phi_{n}|, &&G_n=\alpha_n|\Phi_{n},A\Phi_{n+1},A^3\Phi_{n+1},AT\Phi_{n}|,
\end{eqnarray}
with 
\begin{eqnarray}\label{sec5-fig3-sols3}
    \Phi_n=(\phi_{n}(k_1^+,c_1^+,d_1^+),~ \phi_{n}(k_2^+,c_2^+,d_2^+),~ \psi_{n}(k_1^+,c_1^+,d_1^+),~ \psi_{n}(k_2^+,c_2^+,d_2^+))^T.
\end{eqnarray}
It allows us to get insight into interactions, for example, between two solitons, between soliton and waves and between two waves. Examples are demonstrated in Fig.\ref{fig-classical-diag-sols3} without giving detail expressions.

\captionsetup[figure]{labelfont={bf},name={Fig.},labelsep=period}

When $m=1$ and $\mathbf K_2=J_2(k_1)$ is Jordan matrix,
solution $Q_n=G_n/F_n$ can be investigated by using  the Casoratians \eqref{sec5-sols3-det1} with components
\begin{eqnarray}\label{sec5-classical-jordan4}
    \Phi_n=(\phi_{n}(k_1^+,c_1^+,d_1^+),~ \partial_{k_1}\phi_{n}(k_1^+,c_1^+,d_1^+),~ \psi_{n}(k_1^+,c_1^+,d_1^+),~ \partial_{k_1}\psi_{n}(k_1^+,c_1^+,d_1^+))^T.
\end{eqnarray}

\begin{figure}[ht]
\centering
\subfigure[ ]{
\begin{minipage}[t]{0.32\linewidth}
\centering
\includegraphics[width=2.0in]{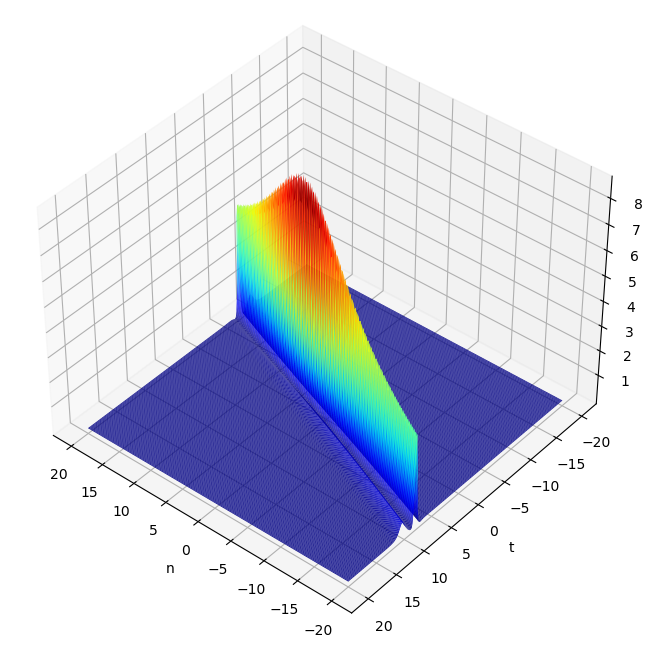}
\end{minipage}%
}%
\subfigure[ ]{
\begin{minipage}[t]{0.32\linewidth}
\centering
\includegraphics[width=2.0in]{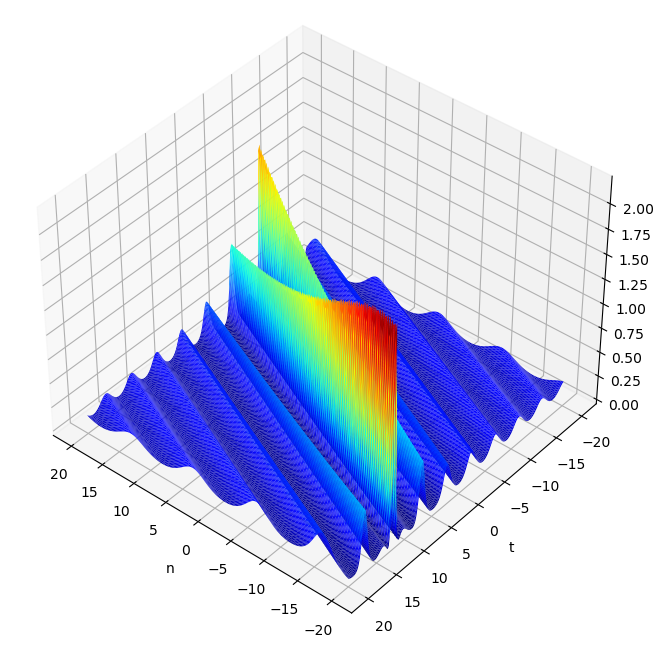}
\end{minipage}%
}%

\caption{\label{fig-classical-jordan-sols1}Shape and motion of second order solutions (Jordan matrix, $|Q_n|^2$ is presented) for the focusing sd-mKdV equation. (a) soliton-soliton-like Interaction with with $a_0=0.2,~ k_1=0.5,~ c_1=d_1=1$.      (b) soliton-wave-like interaction with $a_0=0.4,~ k_1=0.2,~ c_1=d_1=1$. 
}
\end{figure}

It allows us to investigate behaviors similar to interaction between two solitons or interaction between soliton and doubly periodic wave. Examples are shown in Fig.\ref{fig-classical-jordan-sols1} without giving detail expressions. 

\captionsetup[figure]{labelfont={bf},name={Fig.},labelsep=period}

\subsubsection{Rational solutions}
To obtain rational solutions, a new parameter $\kappa$ is introduced, which satisfies
\begin{eqnarray}
    \kappa=\sqrt{(e^k-e^{-k})^2-4 a_0^2}, && e^k=\frac12\sqrt{\kappa^2+4+4 a_0^2}+\frac12\sqrt{\kappa^2+4 a_0^2}.
\end{eqnarray}
The  solution pair $\phi_n(k,c,d)$ and $\psi_n(k,c,d)$ given by formula \eqref{phipsi-sol-pairs1}  can be rewritten as 
\begin{subequations}
    \begin{eqnarray}
        \hat\phi_n(\kappa, c, d)= ce^{\hat\lambda n+\hat\eta t}+ de^{-\hat\lambda n-\hat\eta t},\\
        \hat\psi_n(\kappa, c, d)= -c\hat\xi(\kappa)e^{\hat\lambda n+\hat\eta t}- d\hat\xi(-\kappa)e^{-\hat\lambda n-\hat\eta t},
    \end{eqnarray}
    with
    \begin{eqnarray}
        e^{\hat\lambda}=\frac{\kappa+\sqrt{\kappa^2+4+4 a_0^2}}{\sqrt{4+4 a_0^2}},&\hat\eta=\frac12{\kappa\sqrt{\kappa^2+4+4a_0^2}}, &\hat\xi(\kappa)=\frac{\sqrt{\kappa^2+4 a_0^2}-\kappa}{2a_0}.
    \end{eqnarray}
\end{subequations}

Assume that $d(\kappa)=-c(-\kappa)$. Taking a formal Taylor expansion of $c$ and $d$  results to
\begin{eqnarray}\label{taylor-cd}
    c=\sum_{j=0}^\infty c^{(j)}\kappa^j, && d=\sum_{j=0}^\infty -c^{(j)} (-\kappa)^j.
\end{eqnarray}
It is  easy to verify that  functions $\hat\phi_n(\kappa,c,d)$ and $\hat\phi_n(\kappa,c,d)$ can be formally expanded into Taylor series with only odd order terms of the parameter $\kappa$, i.e.,

\begin{eqnarray}
    \hat \phi_n(\kappa,c,d)=\sum_{j=0}^\infty \phi_{n}^{(2j+1)}\kappa^{2j+1}, && \hat \psi_n(\kappa,c,d)=\sum_{j=0}^\infty \psi_{n}^{(2j+1)}\kappa^{2j+1}.
\end{eqnarray}
Redefining the component vectors with
\begin{subequations}\label{sec5-sd-mKdV-rat1}
    \begin{eqnarray}
        &&\Phi_n=(\phi_{n}^{(1)},\phi_{n}^{(3)},...,\phi_{n}^{(2m+1)},\psi_{n}^{(1)},\psi_{n}^{(3)},...,\psi_{n}^{(2m+1)})^T,\\
        &&\Psi_n=(\psi_{n}^{(1)},\psi_{n}^{(3)},...,\psi_{n}^{(2m+1)},-\phi_{n}^{(1)},-\phi_{n}^{(3)},...,-\phi_{n}^{(2m+1)})^T,
    \end{eqnarray}
\end{subequations}
and reconstructing matrix $A=Diag(e^{K_{m+1}},e^{-K_{m+1}})$ by using
\begin{eqnarray}\label{sec5-sd-mKdV-rat2}
    e^{K_{m+1}}=\begin{bmatrix}
        \zeta_0&0&0&\cdots&0\\
        \zeta_2&\zeta_0&0&\cdots&0\\
        \vdots&\vdots&\vdots&\ddots&\vdots\\
        \zeta_{2m}&\zeta_{2m-2}&\zeta_{2m-4}&\cdots&\zeta_0
    \end{bmatrix}, && \textrm{~~with~~} \zeta_{2j}=\frac{1}{(2j)!}\partial_{\kappa}^{2j}e^k|_{\kappa=0},
\end{eqnarray}
the Casoratians $G_n$ and $F_n$ defined by \eqref{sec3-transform2} yeild rational solution by variable transformation $Q_n=G_n/F_n$.

\begin{figure}[ht]
\centering
\subfigure[ ]{
\begin{minipage}[t]{0.32\linewidth}
\centering
\includegraphics[width=2.0in]{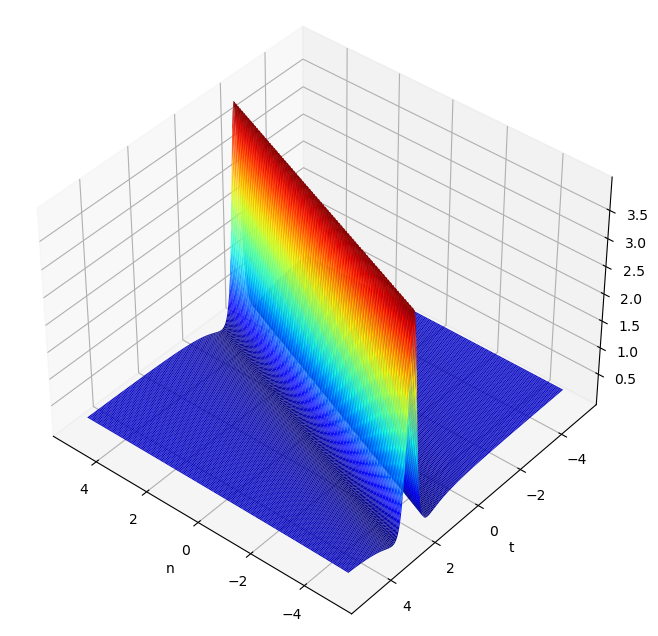}
\end{minipage}\label{fig-classical-rational-sol1}
}%
\subfigure[ ]{
\begin{minipage}[t]{0.32\linewidth}
\centering
\includegraphics[width=2.0in]{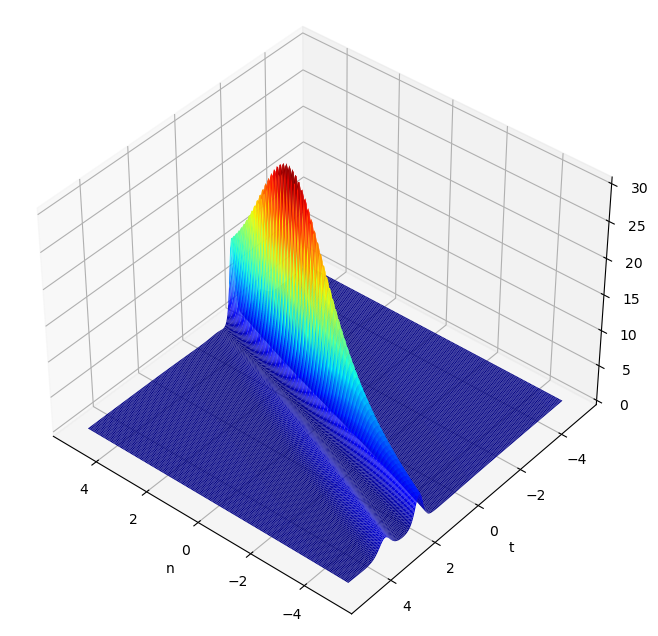}
\end{minipage}\label{fig-classical-rational-sol2}
}%

\caption{\label{fig-classical-rational-sols1}Shape and motion of rational solutions ($|Q_n|^2$ is presented) for the focusing sd-mKdV equation. (a) A first order rational solution with $a_0=0.5$ and $c=1$.    (b) A second order rational solution with $a_0=0.5$ and $c=1$.
}
\end{figure}

When $m=0$, the first order rational solution is explicitly presented by 
\begin{eqnarray}
    Q_n=-a_0(1+\frac{U_n}{W_n})
\end{eqnarray}
with
\begin{eqnarray}
    W_n=\Bigr[({n}+2({1+a_0^2})t+2\frac{c_1}{c_0}\sqrt{1+a_0^2})^2+({n}+2({1+a_0^2})t+2\frac{c_1}{c_0}\sqrt{1+a_0^2}-\frac{\sqrt{1+a_0^2}}{a_0})^2+\frac{\sqrt{1+a_0^2}}{{a_0}}-{1}\Bigr],\\
    U_n=\frac{1+a_0^2+\sqrt{1+a_0^2}}{a_0},
\end{eqnarray}
an example of which is shown in Fig. \ref{fig-classical-rational-sol1}.
A second order rational solution is demonstrated in Fig.\ref{fig-classical-rational-sol2} without giving detail expressions.

\captionsetup[figure]{labelfont={bf},name={Fig.},labelsep=period}

\subsection{The defocusing reverse-space-time complex sd-mKdV equation}
\subsubsection{solitonic and periodic solutions}

When $m=0$ and $\mathbf K_{1}=k_1$, the solution $Q_n=G_n/F_n$ of the defocusing reverse-space-time complex sd-mKdV equation \eqref{sec2-sd-mKdV-equ4} ($\delta=1$) is determined by Casoratians
\begin{subequations}\label{sec5-nonlocal-for1}    
    \begin{eqnarray}
        F_n=-\sqrt{1-a_0^2}\Bigr[A_1e^{2(a_1 n+a_2t)}+B_1e^{-2(a_1 n+a_2 t)}+C_1e^{2i(b_1n+b_2 t)}+D_1e^{-2i(b_1n+b_2t)}\Bigr],\\
        G_n=\sqrt{1-a_0^2}\Bigr[A_2e^{2(a_1 n+a_2t)}+B_2e^{-2(a_1 n+a_2 t)}+C_2e^{2i(b_1n+b_2 t)}+D_2e^{-2i(b_1n+b_2t)}\Bigr],
    \end{eqnarray}
    with 
    \begin{eqnarray}
        A_1=c_1d_1^*(e^{\lambda_1-\lambda_1^*}-\xi(k_1)\xi^*(-k_1)),   && A_2=c_1d_1^*\xi^*(-k_1)(e^{k_1^*}-e^{k_1}e^{\lambda_1-\lambda_1^*}),\\
        B_1=c_1^*d_1(e^{\lambda_1^*-\lambda_1}-\xi^*(k_1)\xi(-k_1)),   && B_2=-c_1^*d_1\xi^*(k_1)(e^{k_1^*}-e^{k_1}e^{\lambda_1^*-\lambda_1}),\\
        C_1=c_1c_1^*(e^{\lambda_1+\lambda_1^*}+\xi(k_1)\xi^*(k_1)),    && C_2=-c_1c_1^*\xi^*(k_1)(e^{k_1^*}-e^{k_1}e^{\lambda_1+\lambda_1^*}),\\
        D_1=d_1d_1^*(e^{-\lambda_1-\lambda_1^*}+\xi(-k_1)\xi^*(-k_1)), && D_2=d_1d_1^*\xi^*(-k_1)(e^{k_1^*}-e^{k_1}e^{-\lambda_1-\lambda_1^*}),\\
        \lambda_1=\lambda(k_1)=a_1+b_1i, && \eta_1=\eta(k_1)=a_2+b_2i.
    \end{eqnarray}    
\end{subequations}
Here, $\lambda,~\eta$ and $\xi$ are defined in \eqref{phipsi-sol-pairs2-1}-\eqref{phipsi-sol-pairs2-3} with $b_0=a_0$ and $a_0\in \mathbb R$. 
Similar to the classical situations, this formula allows us to easily analyze the obtained solution $Q_n$ and there are also two interesting cases.

\begin{figure}[ht]
\centering
\subfigure[ ]{
\begin{minipage}[t]{0.32\linewidth}
\centering
\includegraphics[width=2.0in]{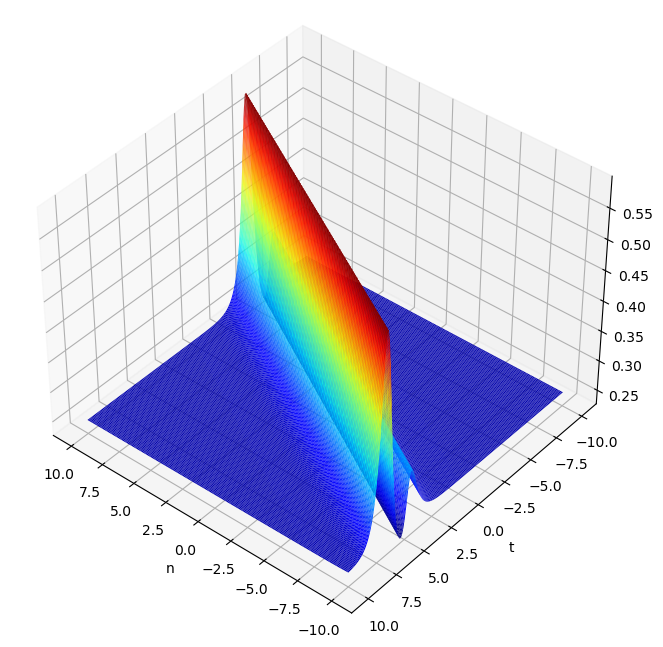}
\end{minipage}\label{fig-nonlocal-single-sol1}
}%
\subfigure[ ]{
\begin{minipage}[t]{0.32\linewidth}
\centering
\includegraphics[width=2.0in]{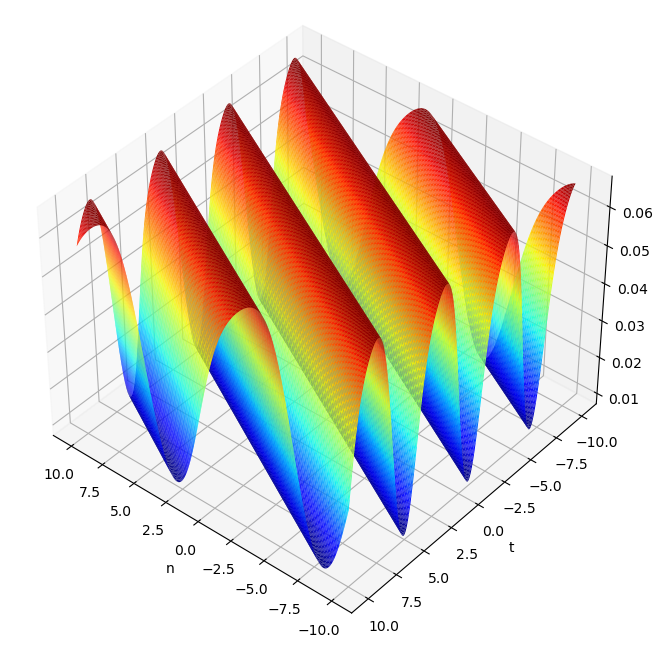}
\end{minipage}\label{fig-nonlocal-single-sol2}
}%

\caption{\label{fig-nonlocal-single-sols1} Shape and motion of first order solutions ($|Q_n|^2$ is presented) for the defocusing reverse-space-time sd-mKdV equation. (a) Soliton  with $a_0=0.5,~ k_1=0.2j,~ c_1=-d_1=1$.    (b) Doubly periodic wave  with $a_0=0.1,~ k_1=0.3j,~ c_1=-d_1=1$. 
}
\end{figure}

The first case is when
\begin{eqnarray}
    k_1\in i\mathbb R, && 4>4a_0^2>-(e^{k_1}-e^{-k_1})^2,
\end{eqnarray}
which yields   $b_1=b_2=0$  and $|\xi(k_1)|=1$. In this case, the solution is reduced to
\begin{eqnarray}
    Q_n=-\frac{A_2e^{X_1}+B_2e^{-X_1}+C_2+D_2}{A_1e^{X_1}+B_1e^{-X_1}+C_1+D_1},&\textrm{with~~} X_1=2(a_1 n+a_2t),
\end{eqnarray}
which is a soliton traveling parallel to $X_1=0$ with a constant velocity $-a_2/a_1$.  For a fixed $t$, it is easy to see that this solution satisfies
\begin{eqnarray}
    \lim_{n\rightarrow\pm\infty}|Q_n|^2=a_0^2,
\end{eqnarray}
which means its squared envelope lives on plane characterized by $a_0^2$. An example is shown in Fig. \ref{fig-nonlocal-single-sol1}.

The second case is when
\begin{eqnarray}
    k_1\in i\mathbb R, & 4a_0^2<-(e^{k_1}-e^{-k_1})^2, &a_0^2<1,
\end{eqnarray}
which yeilds $a_1=a_2=0$.  In this scenario, the solution reduces to

\begin{eqnarray}
    Q_n=-\frac{A_2+B_2+C_2e^{X_1}+D_2e^{-X_1}}{A_1+B_1+C_1e^{X_1}+D_1e^{-X_1}}, &\textrm{with~~} X_1=2i(b_1 n+b_2t),
\end{eqnarray}
which is doubly  periodic with period $T_1=\pi/b_1$ in the $n$-direction and $T_2=\pi/b_2$ in the $t$-direction. An example is shown in Fig.\ref{fig-nonlocal-single-sol2}.

\captionsetup[figure]{labelfont={bf},name={Fig.},labelsep=period}

\begin{figure}[ht]
\centering
\subfigure[ ]{
\begin{minipage}[t]{0.32\linewidth}
\centering
\includegraphics[width=2.0in]{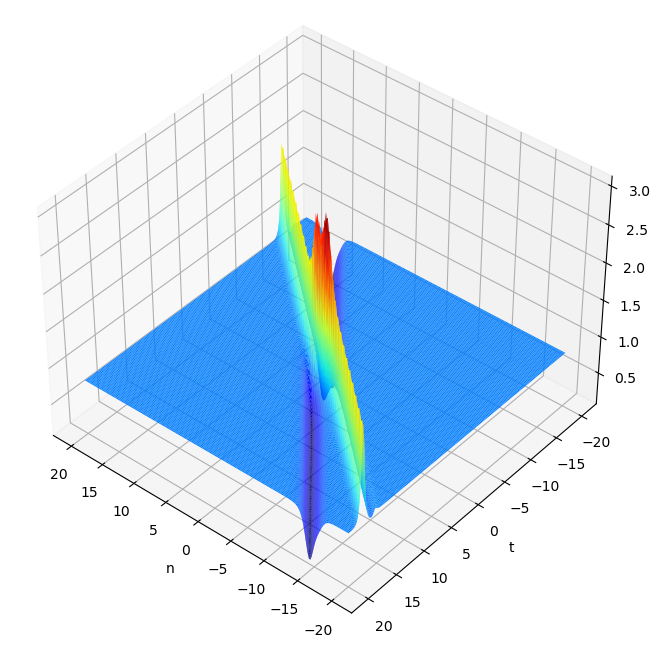}
\end{minipage}%
}%
\subfigure[ ]{
\begin{minipage}[t]{0.32\linewidth}
\centering
\includegraphics[width=2.0in]{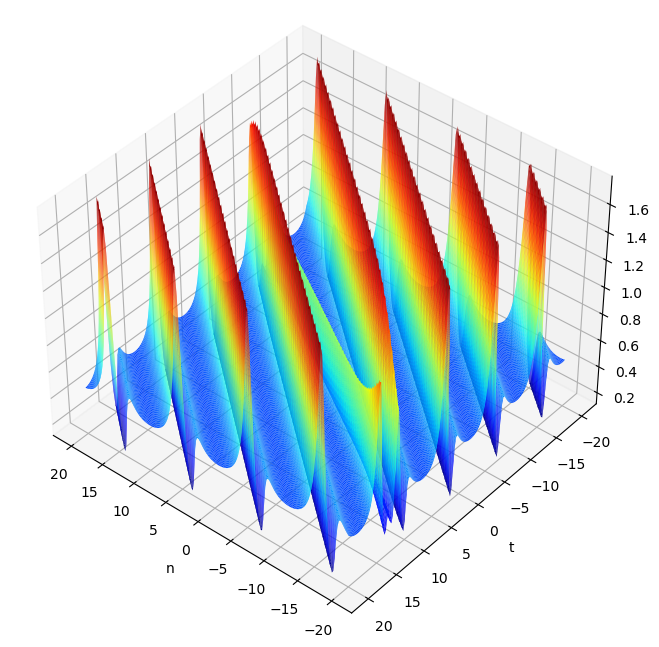}
\end{minipage}%
}%
\subfigure[ ]{
\begin{minipage}[t]{0.32\linewidth}
\centering
\includegraphics[width=2.0in]{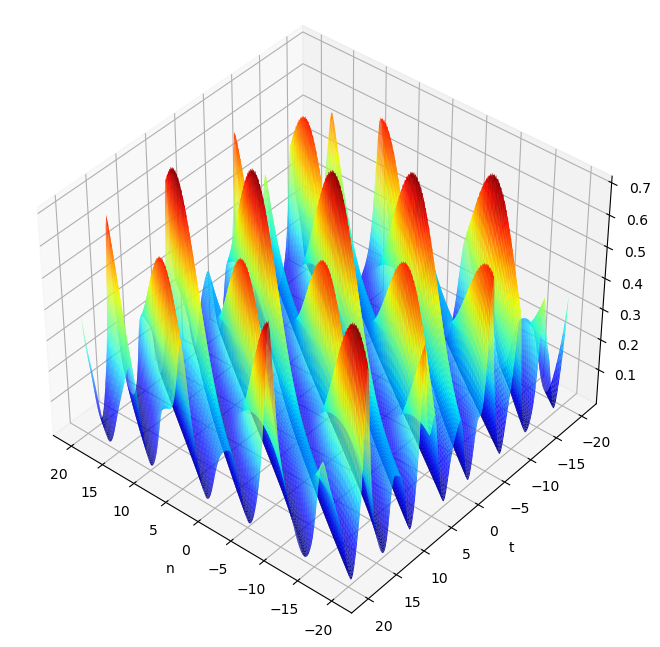}
\end{minipage}%
}%

\caption{\label{fig2-csd-mKdV-sols2}Shape and motion of second order solutions (diagonal matrix, $|Q_n|^2$ is presented) for the defocusing reverse-space-time sd-mKdV equation. (a) Interaction between two solitons  with $a_0=0.9,~ k_1=0.3j, k_2=0.7j~ c_1=d_1=c_2=d_2=1$. (b)  Interaction between soliton and periodic wave  with $a_0=0.6,~ k_1=0.3j, k_2=0.7j~ c_1=d_1=c_2=d_2=1$. (c)  Interaction between two periodic waves  with $a_0=0.1,~ k_1=0.2j, k_2=0.6j~ c_1=d_1=c_2=d_2=1$.
}
\end{figure}

When $m=1$ and $\mathbf K_2=diag(k_1,k_2)$ is diagonal,  solution $Q_n=G_n/F_n$ can be investigated by using  the following Casoratians
\begin{subequations}\label{sec5-nonlocal-sols3}
    \begin{eqnarray}\label{sec5-nonlocal-sols3-m2}
        F_n=\alpha_n|\Phi_{n+1},A^2\Phi_{n+1},T\Phi_{1-n}^*(-t),A^2T\Phi_{1-n}^*(-t)|, &&G_n=\alpha_n|\Phi_{n},A\Phi_{n+1},A^3\Phi_{n+1},AT\Phi_{1-n}^*(-t)|,
    \end{eqnarray}
    with 
    \begin{eqnarray}
        \Phi_n=(\phi_{n}(k_1^+,c_1^+,d_1^+),~ \phi_{n}(k_2^+,c_2^+,d_2^+),~ \psi^*_{1-n,-t}(k_1^+,c_1^+,d_1^+),~ \psi^*_{1-n,-t}(k_2^+,c_2^+,d_2^+))^T,
    \end{eqnarray}
\end{subequations}
which allows us to study interactions between solitons, between periodic wave and soliton and between periodic waves. Examples are demonstrated in  Fig.\ref{fig2-csd-mKdV-sols2} without giving detail expressions.

\captionsetup[figure]{labelfont={bf},name={Fig.},labelsep=period}

\begin{figure}[ht]
\centering
\begin{minipage}[t]{0.32\linewidth}
\centering
\includegraphics[width=2.0in]{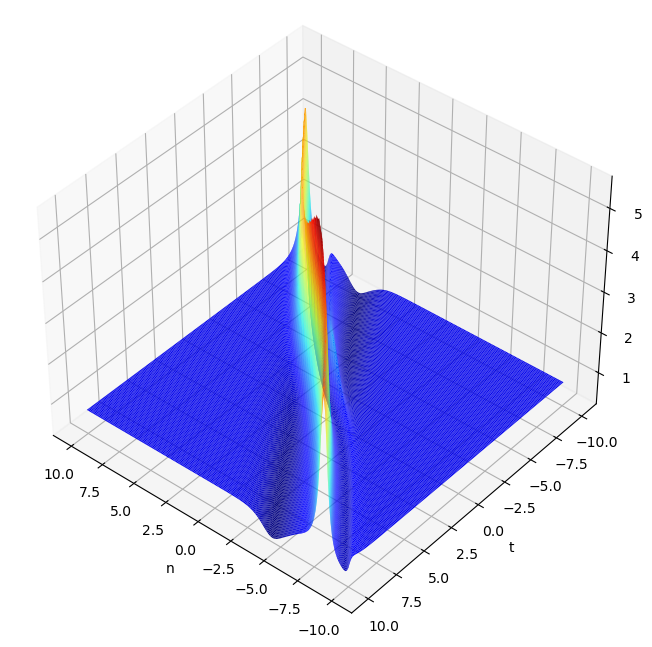}
\end{minipage}%
\caption{\label{fig-nonlocal-jordan-sols1}Shape and motion of a second order solution (Jordan matrix, $|Q_n|^2$ is presented) for the defocusing reverse-space-time sd-mKdV equation with $a_0=0.9,~ k_1=0.8j~ c_1=d_1=1$.
}
\end{figure}

When $m=1$ and $\mathbf K_2=J_2(k_1)$ is a Jordan matrix,
solution $Q_n=G_n/F_n$ can be investigated by using  the formula \eqref{sec5-nonlocal-sols3-m2} with component vector
\begin{eqnarray}\label{sec5-fig4-sols4}
    \Phi_n=(\phi_{n}(k_1^+,c_1^+,d_1^+),~ \partial_{k_1}\phi_{n}(k_1^+,c_1^+,d_1^+),~ \psi^*_{1-n,-t}(k_1^+,c_1^+,d_1^+),~ (\partial_{k_1}\psi_{1-n,-t}(k_1^+,c_1^+,d_1^+))^*)^T,
\end{eqnarray}
which allows us to study, for example, soliton-like interaction. An example is shown in Fig. \ref{fig-nonlocal-jordan-sols1} without giving detail expressions.

\captionsetup[figure]{labelfont={bf},name={Fig.},labelsep=period}

\subsubsection{Rational solutions}

Similar to the classical scenario discussed in Sec.\ref{sec51}, rational solutions for equation \eqref{sec2-sd-mKdV-equ4} ($\delta=1$)  can be investigated by a new parameter $\kappa$ satisfying

\begin{eqnarray}
    \kappa=\sqrt{(e^k-e^{-k})^2+4 a_0^2}, && e^k=\frac12\sqrt{\kappa^2+4-4 a_0^2}+\frac12\sqrt{\kappa^2-4 a_0^2}.
\end{eqnarray}
The  solution pair $\phi_n(k,c,d)$ and $\psi_n(k,c,d)$ given by formula \eqref{phipsi-sol-pairs1}  can be rewritten as 
\begin{subequations}
    \begin{eqnarray}
        \hat\phi_n(\kappa, c, d)= ce^{\hat\lambda n+\hat\eta t}+ de^{-\hat\lambda n-\hat\eta t},\\
        \hat\psi_n(\kappa, c, d)= -c\hat\xi(\kappa)e^{\hat\lambda n+\hat\eta t}- d\hat\xi(-\kappa)e^{-\hat\lambda n-\hat\eta t},
    \end{eqnarray}
    with
    \begin{eqnarray}
        e^{\hat\lambda}=\frac{\kappa+\sqrt{\kappa^2+4-4 a_0^2}}{\sqrt{4-4 a_0^2}},&\hat\eta=\frac12{\kappa\sqrt{\kappa^2+4-4a_0^2}}, &\hat\xi(\kappa)=\frac{\sqrt{\kappa^2-4 a_0^2}-\kappa}{2a_0}.
    \end{eqnarray}
\end{subequations}

\begin{figure}[ht]
\centering
\subfigure[ ]{
\begin{minipage}[t]{0.32\linewidth}
\centering
\includegraphics[width=2.0in]{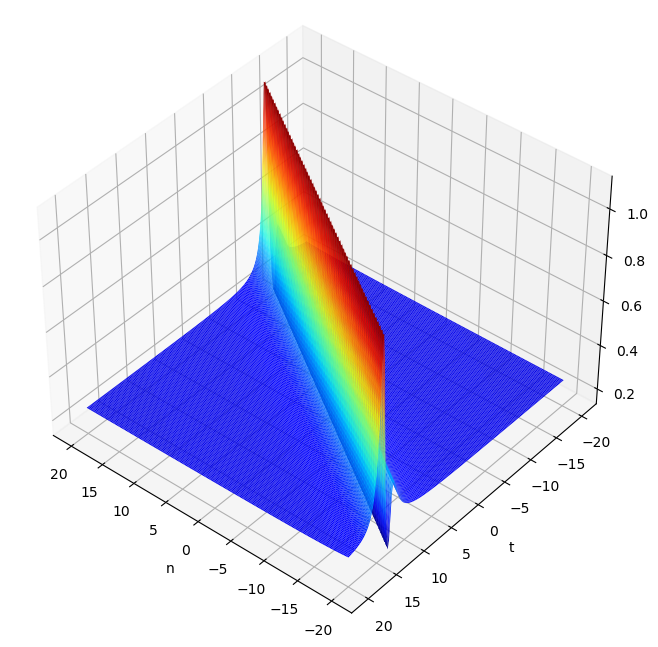}
\end{minipage}%
}%
\subfigure[ ]{
\begin{minipage}[t]{0.32\linewidth}
\centering
\includegraphics[width=2.0in]{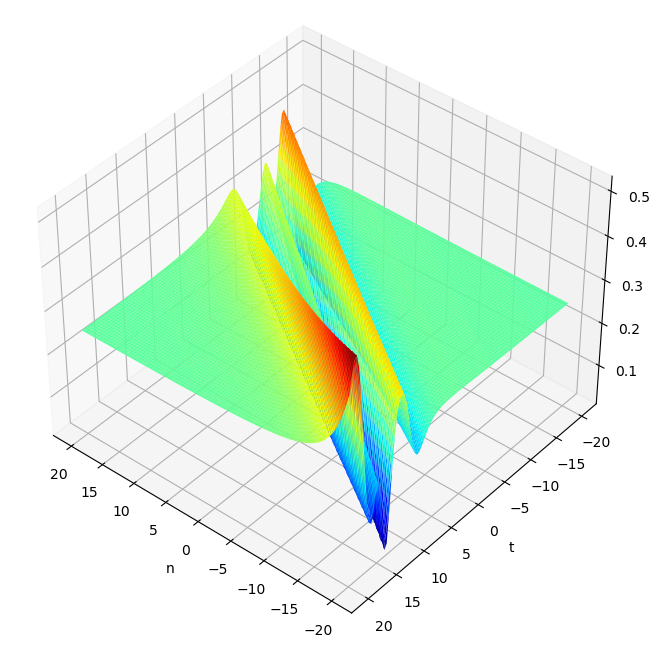}
\end{minipage}%
}%

\caption{\label{fig-nonlocal-rational-sols1}Shape and motion of rational solutions ($|Q_n|^2$ is presented) for the defocusing reverse-space-time sd-mKdV equation.  (a) First order rational solution with $m=0$, $a_0=0.6$ and $c=1$. (b) Second order rational solution with $m=0$, $a_0=0.5$ and $c=1+2\kappa+2i\kappa^2$.
}
\end{figure}

Assume that $d(\kappa)=-c(-\kappa)$, which allows us to take a formal Taylor expansion of $c$ and $d$ and results to eq. \eqref{taylor-cd}.
Functions $\hat\phi_n(\kappa,c,d)$ and $\hat\phi_n(\kappa,c,d)$ can be formally expanded into Taylor series with only odd order terms of the parameter $\kappa$, namely,
\begin{eqnarray}
    \hat\phi_n(\kappa,c,d)=\sum_{j=0}^\infty \phi_{n}^{(2j+1)}\kappa^{2j+1}, && \hat\psi_n(\kappa,c,d)=\sum_{j=0}^\infty \psi_{n}^{(2j+1)}\kappa^{2j+1}.
\end{eqnarray}
Redefining the Casoratian vectors with
\begin{subequations}\label{sec5-nonlocal-rat1}
    \begin{eqnarray}
        &&\Phi_n=(\phi_{n}^{(1)},\phi_{n}^{(3)},...,\phi_{n}^{(2m+1)},\psi_{1-n,-t}^{(1)*},\psi_{1-n,-t}^{(3)*},...,\psi_{1-n,-t}^{(2m+1)*})^T,\\
        &&\Psi_n=(\psi_{n}^{(1)},\psi_{n}^{(3)},...,\psi_{n}^{(2m+1)},-\phi_{1-n,-t}^{(1)*},-\phi_{1-n,-t}^{(3)*},...,-\phi_{1-n,-t}^{(2m+1)*})^T,
    \end{eqnarray}
\end{subequations}
and reconstructing matrix $A=Diag(e^{K_{m+1}},e^{K^*_{m+1}})$ by using
\begin{eqnarray}\label{sec5-nonlocal-rat2}
    e^{K_{m+1}}=\begin{bmatrix}
        \zeta_0&0&0&\cdots&0\\
        \zeta_2&\zeta_0&0&\cdots&0\\
        \vdots&\vdots&\vdots&\ddots&\vdots\\
        \zeta_{2m}&\zeta_{2m-2}&\zeta_{2m-4}&\cdots&\zeta_0
    \end{bmatrix}, && \textrm{~~with~~} \zeta_{2j}=\frac{1}{(2j)!}\partial_{\kappa}^{2j}e^k|_{\kappa=0},
\end{eqnarray}
the Casoratians $G_n$ and $F_n$ defined by \eqref{sec3-phipsi} yeild rational solutions by variable transformation $Q_n=G_n/F_n$.  Two examples of rational solutions are shown in Fig.\ref{fig-nonlocal-rational-sols1} without giving detail expressions.

\captionsetup[figure]{labelfont={bf},name={Fig.},labelsep=period}

\section{Conclusions}
In this paper, the bilinearization-reduction approach \citep{ChenDLZ-SAM2018}  is implemented to obtain the exact solutions with nonzero backgrounds for the sd-mKdV equation, the complex sd-mKdV equation, the  reverse-space-time sd-mKdV equation and the reverse-space-time complex sd-mKdV equation. The first two equations are obtained with local real and complex reductions, while the later two equations are obtained with nonlocal real and complex reductions.

The bilinearization-reduction approach starts from transforming the unreduced system \eqref{sec2-sd-mKdV-coupled1} of the coupled sd-mKdV equations into a bilinear system \eqref{sec3-bilinear1} with nonzero backgrounds, utilizing the variable transformation \eqref{sec3-tra1}. 
This bilinearization admits quasi double Casoratian solutions  \eqref{sec3-transform2} with a pair of solvable component vectors $\Phi_n$ and $\Psi_n$ given by a  matrix system \eqref{sec3-phipsi}.
It turns out that these Casoratian solutions  are also solutions to the unreduced system \eqref{sec2-sd-mKdV-coupled1} through the  variable transformation  \eqref{sec3-tra1}.
When neutralizing nonzero backgrounds, the bilinearization \eqref{sec3-bilinear1}  degenerates to a bilinear formula with zero background,  which has been discussed in \cite{FengZS-IJMPB2020}.

For a plane background $(q_n,~r_n)=(a_0,\delta a_0)$ with $a_0\in\mathbb R$, the vectors $\Phi_n$ and $\Psi_n$ for the unreduced system \eqref{sec2-sd-mKdV-coupled1} and its reductions \eqref{sec2-sd-mKdV-eqs1} are explicitly constructed by  utilizing scalar functions $\phi_n$ and $\psi_n$ defined in \eqref{phipsi-sol-pairs1}, allowing us to obtain Casoratian solutions. These solutions are characterized by the canonical form of matrix $A$, since its similar matrices can lead to the same solutions (see Theorem \ref{sec2-theorem1}). To showcase rich dynamics in these solutions,  solitons, periodic waves and  rational solutions for the focusing sd-mKdV equation and for the defocusing  reverse-space-time complex sd-mKdV equation are demonstrated.

\section*{Acknowledgments}
This article is supported by the NSFC grant (Nos. 12201580, 12071432), Zhejiang Provincial Natural Science Foundation (No. LZ24A010007) and the China Postdoctoral Science Foundation (2022M722921).

The authors would like to thank Kui Chen of Zhejiang Lab for his helpful comments.

Declaration of Interests. The authors report no conflict of interest.

\appendix
\section{Proof of theorem \ref{sec2-theorem1}\label{app1-sec1}}
To prove theorem \ref{sec2-theorem1},  the identity \citep{FreemanN-PRSL1983}
\begin{eqnarray}\label{app1-possion}
|M,a,b||M,c,d|-|M,a,c||M,b,d|+|M,a,d||M,b,c|=0
\end{eqnarray}
is utilized. Here $M$ is an $\alpha\times(\alpha-2)$ matrix,  letters $a, b, c$ and $d$ denote $\alpha$-th order column vectors.

Introducing the following shorthand notation
   \begin{eqnarray}
       |\caso{\beta,\beta+2\mu},\caso{\zeta};\caso{\gamma,\gamma+2\nu},\caso{\xi}|=|A^\beta\Phi_n,A^{\beta+2}\Phi_n,...,A^{\beta+2\mu}\Phi_n, A^{\zeta}\Phi_n;A^\gamma\Phi_n,A^{\gamma+2}\Phi_n,...,A^{\gamma+2\nu}\Psi_n,A^{\xi}\Psi_n|,
   \end{eqnarray}
with parameters $\beta, \gamma, \zeta$ and $\xi$ denoting integers and parameters $\mu$ and $\nu$ denoting positive integers, the Casoratians in Theorem \ref{sec2-theorem1} can be rewritten as 
\begin{subequations}
   \begin{eqnarray}
    G_n=|A|^{-1}(g_n+(-1)^mq_nf_n),  && H_n=|A|^{-1}(h_n+(-1)^mr_nf_n),\\
    F_n=|\caso{1,2m+1};\caso{0,2m}|, && f_n=|\caso{1,2m+1};\caso{2,2m+2}|,\\
    g_n=|\caso{1,2m+3};\caso{2,2m}|, && h_n=|\caso{3,2m+1};\caso{0,2m+2}|,
   \end{eqnarray}
\end{subequations}
 in which the matrix equation \eqref{sec3-phipsi-spectral} has been utilized.
 
A direct calculation yields
\begin{subequations}
    \begin{eqnarray*}
    \alpha_n|A| F_{n+1}&=&|\caso{3,2m+3};\caso{0,2m}|-(-1)^{m}q_nh_n-(-1)^mr_ng_n-q_nr_nf_n,\\
    |A|F_{n-1}&=&f_n,\\
    \alpha_n G_{n+1}&=&|\caso{1,2m+3};\caso{0,2m-2}|+(-1)^mq_n|\caso{1,2m+1};\caso{0,2m-2},\caso{2m+2}|\\
    &&-(-1)^mq_n|\caso{1,2m-1},\caso{2m+3};\caso{0,2m}|+(-1)^m\alpha_nq_nF_n\\
    &&+q_nq_n|\caso{1,2m-1};\caso{0,2m+2}|+(-1)^m\alpha_nq_{n+1}F_n,\\
    G_{n-1}&=&|\caso{-1,2m+1};\caso{2,2m}|+(-1)^mq_{n-1}F_n,\\
    2|A|\partial_tG_{n}&=&2q_{n-1}r_ng_n+2(-1)^mq_{n,t}f_n\\
    &&+|\caso{1,2m+1},\caso{2m+5};\caso{2,2m}|+|\caso{1,2m+3};\caso{0},\caso{4,2m}|\\
    &&-|\caso{-1},\caso{3,2m+3};\caso{2,2m}|-|\caso{1,2m+3};\caso{2,2m-2},\caso{2m+2}|\\
    &&+(-1)^{m}q_n|\caso{1,2m+1};\caso{2,2m},\caso{2m+4}|+(-1)^mq_n|\caso{1,2m+1};\caso{0},\caso{4,2m+2}|\\
    &&-(-1)^mq_n|\caso{-1},\caso{3,2m+1};\caso{2,2m+2}|-(-1)^mq_n|\caso{1,2m-1},\caso{2m+3};\caso{2,2m+2}|\\
    &&-2(-1)^mq_{n-1}|\caso{3,2m+3};\caso{0,2m}|+2(-1)^mq_{n-1}f_n+2q_nq_{n-1}h_n,\\
    2\partial_tF_n&=&|\caso{1,2m-1},\caso{2m+3};\caso{0,2m}|+|\caso{1,2m+1};\caso{-2},\caso{2,2m}|\\
    &&-|\caso{-1},\caso{3,2m+1};\caso{0,2m}|-|\caso{1,2m+1};\caso{0,2m-2},\caso{2m+2}|\\
    &&-2(-1)^mq_n|\caso{1,2m-1};\caso{0,2m+2}|-2(-1)^mr_n|\caso{-1,2m+1};\caso{2,2m}|.
    \end{eqnarray*}
\end{subequations}
Substituting the above calculation into Eq. \eqref{sec3-bilinear-equ1} results to

    \begin{eqnarray*}
&&|A|^2(\alpha_nF_{n+1}F_{n-1}+G_nH_n-F_nF_n)\\
&=&|\caso{3,2m+3};\caso{0,2m}|f_n+g_nh_n-|\caso{3,2m+3};\caso{2,2m+2}|F_n,
\end{eqnarray*}
which vanishes by means of eq.\eqref{app1-possion}.

Similarly, Substituting the above calculation into eq. \eqref{sec3-bilinear-equ2} becomes

\begin{subequations}
    \begin{eqnarray*}
&&2|A|(\alpha_nG_{n+1}F_{n-1}-\alpha_nG_{n-1}F_{n+1}-D_tG_{n}\cdot F_n)=S_1+S_2+(-1)^mq_n(S_3+S_4),
\end{eqnarray*}
\end{subequations}
where
\begin{subequations}
    \begin{eqnarray*}
&&S_1=-2|\caso{-1,2m+1};\caso{2,2m}||\caso{3,2m+3};\caso{0,2m}|-|\caso{1,2m+3};\caso{0},\caso{4,2m}|F_n+|\caso{-1},\caso{3,2m+3};\caso{2,2m}|F_n\\
&&+|\caso{1,2m+1};-2,\caso{2,2m}|g_n-|\caso{-1},\caso{3,2m+1};\caso{0,2m}|g_n,\\
&&S_2=2|\caso{1,2m+3};\caso{0,2m-2}|f_n-|\caso{1,2m+1},\caso{2m+5};\caso{2,2m}|F_n+|\caso{1,2m+3};\caso{2,2m-2},\caso{2m+2}|F_n\\
&&+|\caso{1,2m-1},\caso{2m+3};\caso{0,2m}|g_n-|\caso{1,2m+1};\caso{0,2m-2},\caso{2m+2}|g_n,\\
&&S_3=2|\caso{-1,2m+1};\caso{2,2m}|h_n-|\caso{1,2m+1};\caso{0},\caso{4,2m+2}|F_n+|\caso{-1},\caso{3,2m+1};\caso{2,2m+2}|F_n\\
&&+|\caso{1,2m+1};\caso{-2},\caso{2,2m}|f_n-|\caso{-1},\caso{3,2m+1};\caso{0,2m}|f_n,\\
&&S_4=-2|\caso{1,2m-1};\caso{0,2m+2}|g_n-|\caso{1,2m+1};\caso{2,2m},\caso{2m+4}|F_n+|\caso{1,2m-1},\caso{2m+3};\caso{2,2m+2}|F_n\\
&&+|\caso{1,2m+1};\caso{0,2m-2},\caso{2m+2}|f_n-|\caso{1,2m-1},\caso{2m+3};\caso{0,2m}|f_n.
    \end{eqnarray*}
\end{subequations}
Each $S_j$ vanishes by using the identity \eqref{app1-possion} twice,
thus  eq. \eqref{sec3-bilinear-equ2} holds. Eq. \eqref{sec3-bilinear-equ3} can be proved in a similar way.

Suppose matrices $A$ and $J$ are similar through $A=P^{-1}JP$, we can introduce $\Phi'_n=P\Phi_n$ and $\Psi'_n=P\Psi_n$, which again satisfy matrix equations \eqref{sec3-phipsi} with matrix $J$ replacing matrix $A$. The double Casoratians satisfy $F(J,\Phi'_n,\Psi'_n)=|P|F(A,\Phi_n,\Psi_n)$, $G(J,\Phi'_n,\Psi'_n)=|P|G(A,\Phi_n,\Psi_n)$ and $H(J,\Phi'_n,\Psi'_n)=|P|H(A,\Phi_n,\Psi_n)$. This imply that matrix $A$, together with its associated vectors $\Phi_n$ and $\Psi_n$, and its similar matrices, together with their associated vectors, lead to same $Q_n$ and $R_n$.










\bibliographystyle{cas-model2-names}

\bibliography{cas-refs-sd-mKdV}



\end{document}